\documentclass[aps,prl,showpacs,superscriptaddress,twocolumn,longbibliography]{revtex4-1}

\usepackage{amsfonts}
\usepackage{subfigure}
\usepackage{amsmath}
\usepackage{amssymb}
\usepackage{amsbsy} 
\usepackage{epsfig}
\usepackage{graphicx}
\usepackage{epstopdf}

\def\be{\begin{equation}} \def\ee{\end{equation}}
\def\bea{\begin{eqnarray}} \def\eea{\end{eqnarray}}
\def\nn{\nonumber}

\def\lw{\leftarrow}
\def\rw{\rightarrow}

\def\ra{\rangle}
\def\la{\langle}

\begin{document}

\title{Simple formulas of directional amplification from non-Bloch band theory}

\author{Wen-Tan Xue}
\affiliation{Institute for Advanced Study, Tsinghua University, Beijing, 100084, China}

\author{Ming-Rui Li}
\affiliation{Institute for Advanced Study, Tsinghua University, Beijing, 100084, China}

\author{Yu-Min Hu}
\affiliation{Institute for Advanced Study, Tsinghua University, Beijing, 100084, China}

\author{Fei Song}
\affiliation{Institute for Advanced Study, Tsinghua University, Beijing, 100084, China}

\author{Zhong Wang} \altaffiliation{ wangzhongemail@gmail.com }
\affiliation{Institute for Advanced Study, Tsinghua University, Beijing, 100084, China}


\begin{abstract}

Green's functions are fundamental quantities that determine the linear responses of physical systems. The recent developments of non-Hermitian systems, therefore, call for Green's function formulas of non-Hermitian bands. This task is complicated by the high sensitivity of energy spectrums to boundary conditions, which invalidates the straightforward generalization of Hermitian formulas. Here, based on the non-Bloch band theory, we obtain simple Green's function formulas of general one-dimensional non-Hermitian bands. Furthermore, in the large-size limit, these formulas dramatically reduce to finding the roots of a simple algebraic equation. As an application, our formulation provides the desirable formulas for the defining quantities, the gain and directionality, of directional amplification. Thus, our formulas provide an efficient guide for designing directional amplifiers.

\end{abstract}

\maketitle

The responses of a physical system are generally proportional to a small external perturbation, which is captured by the Green's functions. Whereas their explicit formulas are well known for Hermitian energy bands, recent progresses in non-Hermitian systems call for their generalizations. This seemingly straightforward task is hindered by the non-Hermitian skin effect\cite{yao2018edge,kunst2018biorthogonal,lee2018anatomy,alvarez2017,Xiao2019NHSE,Helbig2019NHSE,Ghatak2019NHSE}, meaning the exponential localization of most eigenstates to the boundaries. This effect causes a high sensitivity of Green's functions to the boundary condition, invalidating a straightforward extension of Hermitian formulas. It is the purpose of this paper to obtain general formulas of non-Hermitian Green's functions.

\begin{figure}
\includegraphics[width=8.5cm, height=2.8cm]{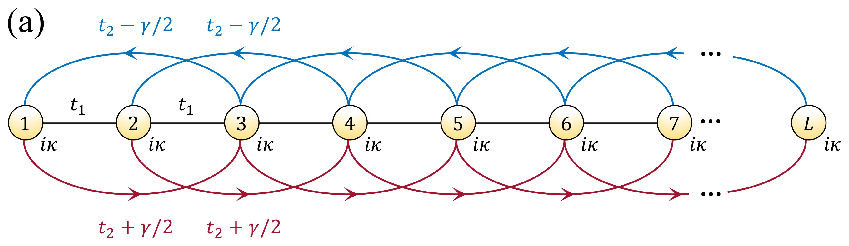}
\includegraphics[width=4.1cm, height=3.8cm]{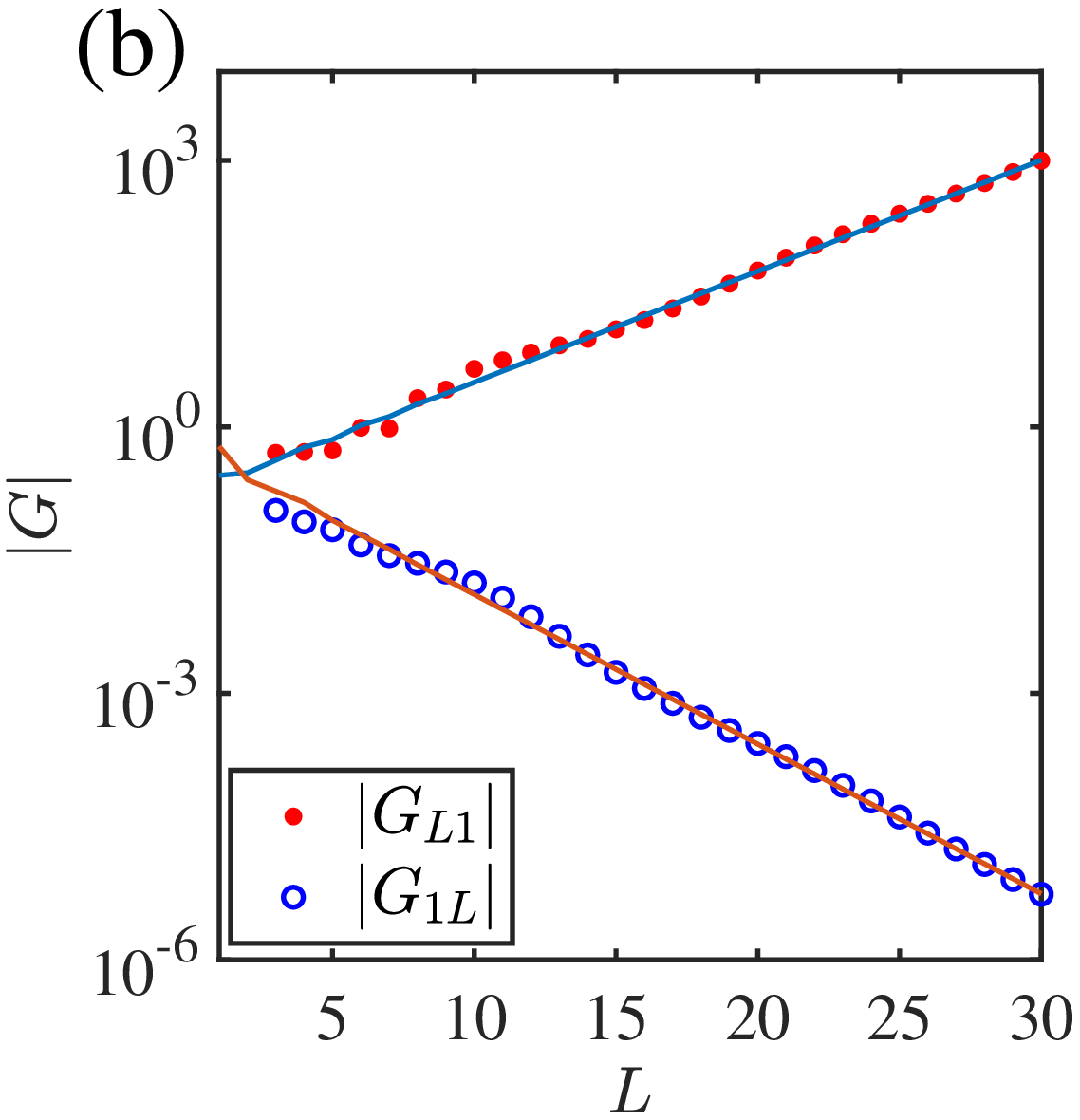}
\includegraphics[width=4.1cm, height=3.8cm]{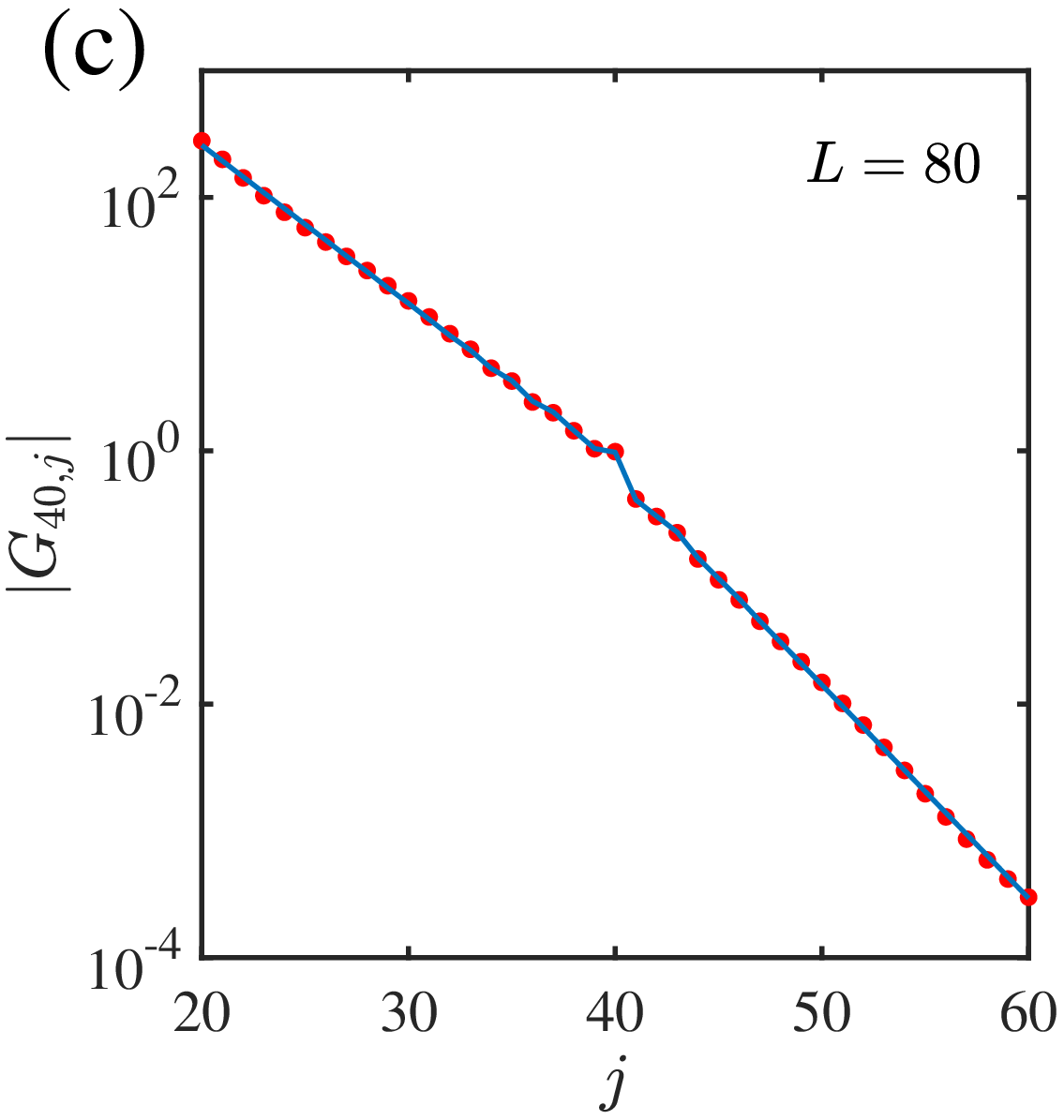}
\includegraphics[width=8.5cm, height=2.8cm]{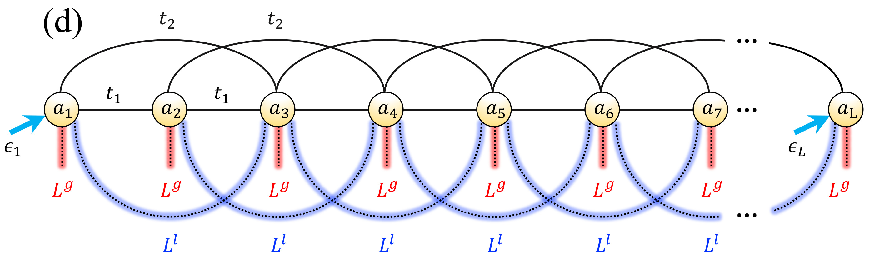}
\caption{  ({\bf a}) The Hamiltonian $H$, with open-boundary condition (OBC) at the two ends. ({\bf b}) $|G_{L1}|$ and $|G_{1L}|$. The corresponding curve represents Eq. (\ref{formula1}) and Eq. (\ref{formula2}), with $|K|=0.292$ and $0.619$, respectively. ({\bf c}) $|G_{40,j}|$ for $L=80$ (dots). Blue curve represents Eq. (\ref{formula}).  For (b)(c), parameters are $t_1=t_2=1$, $\gamma$=4/3, $\kappa=-0.8$, and $\omega=-1.7$.  ({\bf d}) An open quantum system whose effective Hamiltonian is (a). Gain and loss are denoted by $L^{g,l}$ and the external signals by $\epsilon_i$.  \label{fig1} }
\end{figure}

The problem can be simply phrased. Let us consider a general non-Hermitian Hamiltonian $H$ of one-dimensional (1D) lattice with length $L$, with translational symmetry $H_{ij}=H_{i+1,j+1}$ (for $i,j=1,2,\cdots,L-1$). For example, if we take $H_{i,i\pm 1}=t_1, H_{i,i\pm 2}=t_2\mp\gamma/2, H_{ii}=i\kappa$ and all other matrix elements zero, $H$ can be shown pictorially as Fig. \ref{fig1}(a). We take open-boundary condition (OBC) at the two ends\footnote{Our results will be generalizable to other boundary conditions such as domain wall systems. In mathematical literature, OBC corresponds to Toeplitz matrices, while PBC corresponds to circulant matrices. }. Our goal is to find explicit formulas for the frequency-domain Green's function matrix \bea G(\omega)=\frac{1}{\omega -H}. \eea
Although Green's functions have recently been studied to extract non-Hermitian topology\cite{Borgnia2019,Zirnstein2019,Silveirinha2019}, their general and explicit formulas have been lacking. As we will see, this seemingly trivial goal is difficult, if not impossible, to achieve from the standard Brillouin zone (BZ) and Bloch-band framework. Here, we will obtain the $G$ matrix from the non-Bloch band theory\cite{yao2018edge,Yokomizo2019}, which is based on the generalized Brillouin zone (GBZ) originally introduced to understand non-Hermitian topology\cite{yao2018edge,yao2018chern,Yokomizo2019,lee2018anatomy,Longhi2019Probing,
Yang2019Auxiliary,Deng2019,Longhi2020chiral,Kawabata2020nonBloch,liu2019second, Song2019real,Lee2020Unraveling,Yi2020,Bergholtz2020RMP,Ashida2020}.  We obtain a simple integral formula for all the matrix elements $G_{ij}(\omega)$. In particular, for the end-to-end Green's functions, $G_{L1}$ and $G_{1L}$, our integral formula reduces in the large-$L$ limit to: \bea
G_{L1}(\omega)\sim [\beta_M(\omega)]^L,  \quad G_{1L}(\omega)\sim [\beta_{M+1}(\omega)]^{-L}, \label{residue}
\eea where $\beta_{j=1,\cdots,2M}$ are the roots of $h(\beta)=\omega$ ordered as $|\beta_1|\leq \cdots\leq|\beta_{2M}|$. Here, $h(\beta)$ denotes the Bloch Hamiltonian of $H$, under the notation $\beta\equiv e^{ik}$, which takes the general form of  $h(\beta)=\sum_{n=-M}^M h_n\beta^n$ with coefficients $h_n=H_{i,i+n}$, $M$ being the hopping range. For example, for the model Fig. \ref{fig1}(a), we have $M=2$ and \bea
h(\beta)=&&(t_2+\frac{\gamma}{2})\beta^{-2} + t_1\beta^{-1}  + i\kappa +t_1\beta  +(t_2-\frac{\gamma}{2})\beta^2.\quad \label{bloch}
\eea
As we will show, the presence in Eq. (\ref{residue}) of the \emph{middle two} roots, namely the $M$-th and $(M+1)$-th of the $2M$ roots, reflects the GBZ origin of Eq. (\ref{residue}).

Among various applications, our formulas are important for directional amplifiers (or nonreciprocal amplifiers). In such devices, signals are amplified in a preferred direction and suppressed in the reversed direction, which protects the signal sources; such feature is essential to a wide range of applications in classical and quantum information processing\cite{Abdo2013directional,Abdo2014,Sliwa2015Reconfigurable,jalas2013what,feng2011nonreciprocal,Caloz2018, fleury2014sound,sounas2017non,yu2009complete,estep2014magnetic,soljacic2003nonlinear,Metelmann2015,ranzani2015graph, fang2017generalized,barzanjeh2017mechanical,Peterson2017,bernier2017nonreciprocal,xu2019nonreciprocal}. Irrespective of device details, their dynamics is generated by effective non-Hermitian Hamiltonians\cite{Metelmann2015,ranzani2015graph,Porras2019, Peterson2017,bernier2017nonreciprocal,fang2017generalized,barzanjeh2017mechanical, xu2019nonreciprocal,ruesink2016nonreciprocity,Barzanjeh2018,McDonald2018phase,Wanjura2019,Wang2019Nonreciprocity,Malz2018,Lepinay2019}, and the gain and directionality are given by the Green's function\cite{Metelmann2015,ranzani2015graph,Porras2019}. To be precise, directional amplification occurs when $|G_{ij}(\omega)|\gg1$ while $|G_{ji}(\omega)|\ll 1$ for a certain pair $(i,j)$, meaning that a $\omega$-frequency signal is amplified from $j$ to $i$, while the back-propagation from $i$ to $j$ is suppressed. Although brute force calculation of Green's function is viable for few-mode cases, it becomes inconvenient for many-mode amplifiers taking the shape of a 1D chain; such 1D amplifiers have the advantage of unlimited gain-bandwidth product without fine tuning\cite{McDonald2018phase,Wanjura2019}. Our Green's function formulas tell their gain and directionality in a simple fashion.

\begin{figure*}
\includegraphics[width=4.2cm, height=4cm]{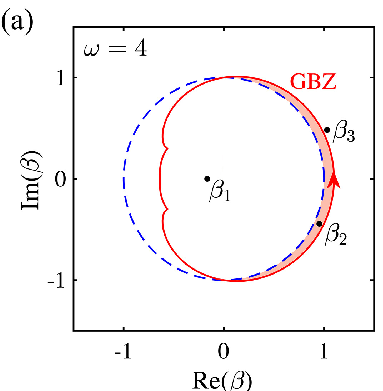}
\includegraphics[width=5cm, height=4cm]{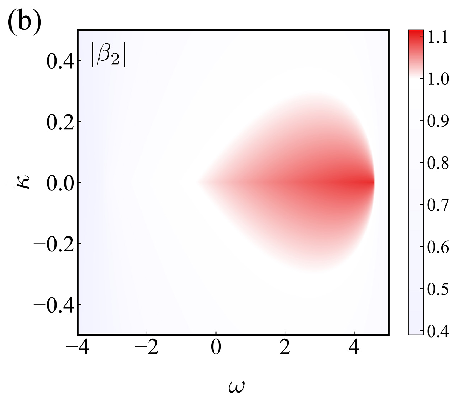}
\includegraphics[width=4.2cm, height=4cm]{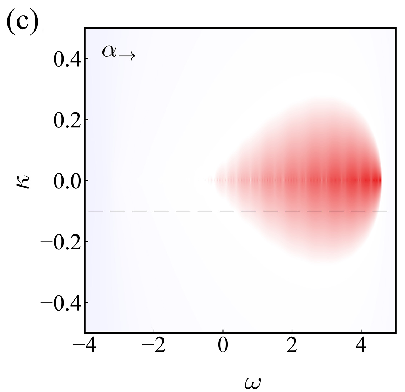}
\includegraphics[width=4.2cm, height=4cm]{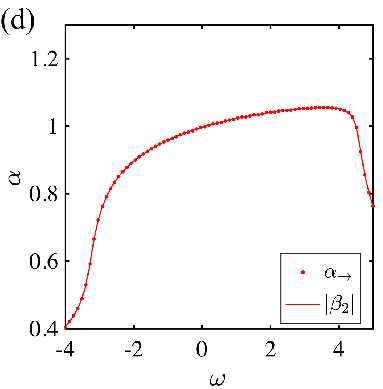}
\includegraphics[width=4.2cm, height=4cm]{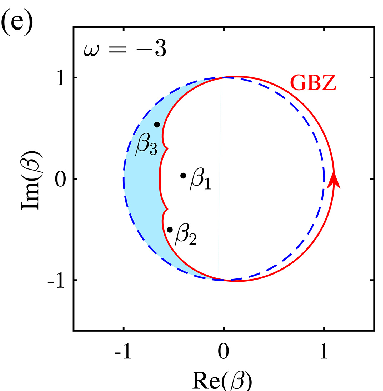}
\includegraphics[width=5cm, height=4cm]{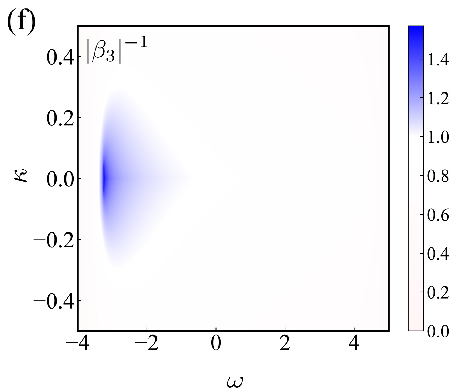}
\includegraphics[width=4.2cm, height=4cm]{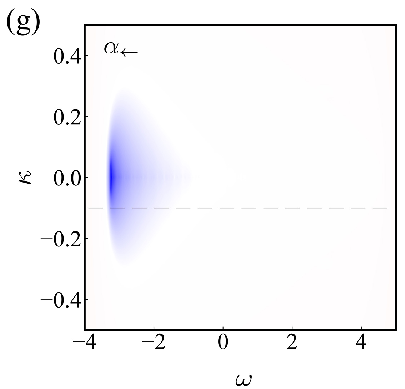}
\includegraphics[width=4.2cm, height=4cm]{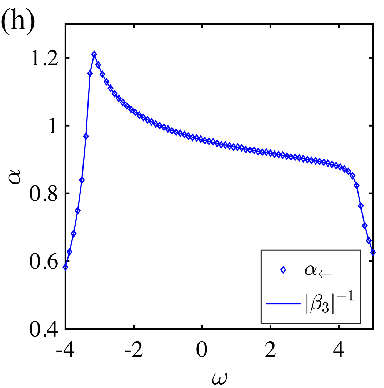}
\caption{ ({\bf a}) GBZ (red solid loop) and BZ (blue dashed circle). $\beta_{1,2,3}$ are roots of $h(\beta)=\omega$ for $\kappa=-0.1$ and $\omega=4$ ($\beta_4$ is outside this region). When there exists a root in the colored region inside GBZ but outside BZ, rightward amplification occurs. ({\bf b}) $|\beta_2|$ as a function of $\kappa,\omega$. ({\bf c}) $\alpha_\rw$ from Eq. (\ref{alpha}). ({\bf d}) $\alpha_\rw$ and $|\beta_2|$ along the cut $\kappa=-0.1$ [dashed line in (c)].
({\bf e}) The same as (a) except that $\omega=-3$. When there exists a root in the colored region inside BZ but outside GBZ, leftward amplification occurs. ({\bf f}) $|\beta_3|^{-1}$. ({\bf g}) $\alpha_\lw$. ({\bf h}) $\alpha_\lw$ and $|\beta_3|^{-1}$ along the cut $\kappa=-0.1$.  Parameter values are $t_1=2$, $t_2=0.3$, $\gamma=0.3$.  }
\label{fig2}
\end{figure*}

\emph{Integral formulas of Green's function.--}Numerically, $G_{ij}(\omega)$ follows an exponential law with respect to $|i-j|$. For example, the end-to-end Green's functions for our specific model [Fig. \ref{fig1}(a)] have the following large-$L$ behaviors,  \bea |G_{L1}(\omega)|\sim (\alpha_{\to})^L, \quad |G_{1L}(\omega)|\sim (\alpha_{\leftarrow})^L, \label{alpha} \eea which is displayed in Fig. \ref{fig1}(b). Knowing $\alpha_\rightarrow$ and $\alpha_\leftarrow$ is important to understanding and designing directional amplification. The condition for rightward amplification, $|G_{L1}|\gg 1$ and $|G_{1L}|\ll 1$, is to require $\alpha_\rw>1$ and $\alpha_\lw<1$; similarly, the condition for leftward amplification is to require $\alpha_\rw<1, \alpha_\lw>1$. Remarkably, such 1D amplification does not suffer from the standard limitation of gain-bandwidth product, because large gain is possible for large $L$, while the bandwidth is independent of $L$\cite{McDonald2018phase,Wanjura2019}.

The values of $\alpha_\rightarrow$ and $\alpha_\leftarrow$ can be derived from the general formulas of $G_{ij}(\omega)$ to be obtained below. To derive the general $G_{ij}(\omega)$, a plausible starting point is the spectral representation $(\omega-H)^{-1}=\sum_n (\omega-E_n)^{-1}|\psi_{nR}\rangle\langle\psi_{nL}|$,
where $E_n$ and $|\psi_{nR(L)}\rangle$ are the eigenvalues and normalized right (left) eigenvectors of $H$ under OBC, namely, $H|\psi_{nR}\ra=E_n|\psi_{nR}\ra, \la\psi_{nL}|H=\la\psi_{nL}|E_n$. Moreover, it is tempting to switch to the BZ and conjecture that  \bea G_{ij}(\omega)=\int_0^{2\pi} \frac{dk}{2\pi} \frac{e^{ik(i-j)}}{\omega-h(k)}.  \eea
With the notation $\beta=e^{ik}$, BZ is the unit circle and the integral becomes: \bea G_{ij}(\omega)=\int_{|\beta|=1}\frac{d\beta}{2\pi i\beta} \frac{\beta^{i-j}}{\omega-h(\beta)}. \label{plausible} \eea
An immediate difficulty is seen after using the residue theorem, which leads to $G_{L1}\sim (\beta_a)^L$, $\beta_a$ being the largest-modulus root of $\omega-h(\beta)=0$ inside the unit circle. This would always imply $\alpha_\rw=|\beta_a|<1$ and forbids any directional amplification. Similarly, one would have $\alpha_\lw=1/|\beta_{a'}|<1$, $\beta_{a'}$ being the smallest-modulus root outside the unit circle. In fact, Eq. (\ref{plausible}) is generally valid only in Hermitian cases, as will become clear below.

The problem with Eq. (\ref{plausible}) is the assumption of the validity of Bloch band theory. In fact, a unique non-Hermitian phenomenon is that, for a broad class of non-Hermitian Hamiltonians, all the eigenstates are localized at the boundaries, which is known as the non-Hermitian skin effect\cite{yao2018edge,kunst2018biorthogonal,lee2018anatomy,alvarez2017,Xiao2019NHSE,Helbig2019NHSE,Ghatak2019NHSE}. This effect suggests that we should remove the usual Bloch-band restriction $|\beta|=1$. Indeed, it has been found that when $\beta$ varies in a closed curve known as the GBZ in the complex plane, the trajectory of $h(\beta)$ is exactly the OBC energy band\cite{yao2018edge,Yokomizo2019}. Note that if $\beta$ varies in the BZ ($|\beta|=1$), the $h(\beta)$ trajectory is the periodic-boundary-condition (PBC) energy band, which is generally different from the OBC energy band. In Hermitian cases, GBZ reduces to the BZ, being consistent with the fact that PBC and OBC bands are the same. The equation that determines the GBZ was found in Refs. \cite{yao2018edge,Yokomizo2019}; we recall their final result below without reproducing the technical derivations. For a hopping range $M$, $h(\beta)=E$ is a $2M$-th order equation with roots $\beta_1(E),\beta_2(E),\cdots,\beta_{2M}(E)$, which are ordered as $|\beta_1|\leq|\beta_2|\leq\cdots\leq|\beta_{2M}|$.
The GBZ equation reads\cite{yao2018edge,Yokomizo2019} \bea |\beta_M(E)|=|\beta_{M+1}(E)|,  \label{GBZ} \eea which is essentially a single-variable equation because $\beta_M$, $\beta_{M+1}$, $E$ are related by $h(\beta_M)=h(\beta_{M+1})=E$. The $\beta_M$ and $\beta_{M+1}$ solutions form a closed loop in the complex plane, which is the GBZ, and the $E$ solutions form the OBC energy bands. Examples of GBZ are shown in Fig. \ref{fig2}(a)(e).


In view of the failure of BZ [e.g. invalidity of Eq. (\ref{plausible})], we propose the following GBZ-based integral formula for $G_{ij}$:
\bea
G_{ij}(\omega)=\int_{\text{GBZ}}\frac{d\beta}{2\pi i\beta} \frac{\beta^{i-j}}{\omega-h(\beta)}, \label{formula}
\eea
which is a main result of this paper. Its proof is provided in Supplemental Material. In practice, this integral is highly convenient to evaluate by the residue theorem, which reduces it to a sum at several roots of $h(\beta)=\omega$. This simplification is enabled by the vital fact that the GBZ is always a closed loop\cite{yao2018edge,Yokomizo2019,Yang2019Auxiliary}. A numerical confirmation of Eq. (\ref{formula}) is shown in Fig. \ref{fig1}(c). Irrespective of $|i-j|$ being large or small, the formula is always precise for $i,j$ not too close to the two ends. At the two ends, because of the boundary effect, a factor $K$ of order unity has to be included:
\bea
G_{L1}(\omega)&=&K\int_{\text{GBZ}}\frac{d\beta}{2\pi i\beta} \frac{\beta^{L-1}}{\omega-h(\beta)},  \label{formula1} \\
G_{1L}(\omega)&=&K\int_{\text{GBZ}}\frac{d\beta}{2\pi i\beta} \frac{\beta^{-(L-1)}}{\omega-h(\beta)}. \label{formula2}
\eea

Now Eq. (\ref{residue}) can be derived as follows. Ordering the roots of $h(\beta)=\omega$ as $|\beta_1|\leq|\beta_2|\leq\cdots\leq|\beta_{2M}|$, one can prove that $\beta_1,\cdots,\beta_M$ are enclosed by the GBZ, while $\beta_{M+1},\cdots,\beta_{2M}$ are not. To see this, suppose that we vary $\omega$ in the complex plane (though $\omega$ is real-valued for physical applications). As long as $\omega$ stays away from the OBC energy spectrum $E_\text{OBC}$, the roots $\beta_i$'s cannot touch the GBZ because GBZ generates $E_\text{OBC}$. Therefore, the number of roots enclosed by GBZ is independent of $\omega$. To determine this number, we consider the $|\omega|\rw\infty$ limit, in which either the $\beta^M$ or $\beta^{-M}$ term dominates $h(\beta)$ and there are $M$ roots with $|\beta|\sim|\omega|^{1/M}\rw\infty$, and also $M$ roots with $|\beta|\sim|\omega|^{-1/M}\rw 0$. Therefore, for any $\omega$, there are $M$ roots $\beta_{1,2,\cdots,M}$ inside the GBZ (For a more rigorous proof, see Refs.\cite{Zhang2020correspondence,Okuma2020}). Now Eq. (\ref{formula1}) and Eq. (\ref{formula2}) can be simplified by the residue theorem. For large $L$, we obtain Eq. (\ref{residue}),
in other words,
\bea
\alpha_\rw =|\beta_M(\omega)|,\quad \alpha_\lw=|\beta_{M+1}(\omega)|^{-1}. \label{residue2}
\eea
Therefore, the \emph{middle two} roots of $h(\beta)=\omega$, $\beta_M$ and $\beta_{M+1}$, determine the gain and directionality, leading to a surprising simplification. The indices $M$ and $M+1$ are difficult to understand from the BZ; they reflect the GBZ origin of Eqs. (\ref{residue})(\ref{residue2}). Eqs. (\ref{formula}-\ref{residue2}) are the central results of this work. For multi-band systems, $h(\beta)$ is a matrix, and Eq. (\ref{residue2}) remains applicable with $\beta_{j=1,\cdots,2M}$ denoting the roots of $\det[\omega-h(\beta)]=0$\cite{supplemental}.

As an application to the model in Fig. \ref{fig1}(a), we show in Fig. \ref{fig2} a quantitative comparison of our theory with the extensive numerical results. For all the parameters investigated, the numerical $\alpha_\rw$ and $\alpha_\lw$ are in excellent agreement with Eq. (\ref{residue2}) with $M=2$. Pictorially, rightward and leftward amplification occurs when a root locates in the colored area in Fig. \ref{fig2}(a) and (e), respectively. Notably, the amplifier can selectively amplify signals in a frequency-dependent direction. In our theory, this frequency-dependent directionality is possible when the GBZ has intersections with the BZ [Fig. \ref{fig2}(a)(e)]. This picture provides a mechanism for designing devices that efficiently integrate directional amplifiers and frequency filters or splitters. On the other hand, the simpler cases of rightward (leftward) unidirectional amplification within the entire bandwidth are realized when the GBZ is entirely outside (inside) the BZ (an example is shown in Supplemental Material).

Intuitively, the nonreciprocal hoppings seem to favor motion towards a preferred direction, causing directional amplification in that direction; e.g. when $|t_2+\gamma/2|>|t_2-\gamma/2|$ in Fig. \ref{fig1}(a), it seems that the directional amplification should be rightward. However, as has been shown above, leftward directional amplification is also seen in certain frequency window. Thus, the simple intuition based on hopping direction fails even qualitatively by telling a wrong amplification direction. Our formulas of $G$ are therefore not merely a matter of quantitative precision, but also important for a qualitative prediction.

Very recently, effort has been made to find OBC Green's function formulas from conventional Brillouin zone\cite{Wanjura2019}. Their results are applicable only to the simplest case where the hoppings exist only between nearest neighbors, in which case our results are consistent with theirs. It is highly challenging, if not impossible, to generalize their approach to the cases beyond nearest-neighbor hopping [e.g. our model Fig. \ref{fig1}(a)]. In contrast, our formulas are straightforward to use for general hoppings. Note that the necessity of taking OBC has been emphasized in recent insightful papers\cite{McDonald2018phase,Wanjura2019}, though general formulas were lacking. In fact, if one takes PBC, the directional amplifiers become dynamically unstable, which results from the endless amplification during the cyclic directional motion.

We note that our results are generalizable to more complicated boundary conditions such as domain wall geometries, for which GBZ remains definable\cite{Deng2019}.  Moreover, as the concept of GBZ also applies to higher dimensions\cite{yao2018chern,liu2019second}, we expect that higher-dimensional generalizations of our formulas remain valid, though their applications rely on efficient algorithms of GBZ, which are called for in higher dimensions.

\emph{Realization in open quantum systems.--}So far, the non-Hermitian Hamiltonian is taken for granted. While it is without question in classical platforms, for example, the Green's function is directly measurable in the topolectrical circuits\cite{Helbig2019NHSE}, we emphasize that our formulas are also applicable to various open quantum systems. For example, we may consider a 1D lattice of coupled bosonic modes, which can be realized in various realistic systems such as optomechanical cavities\cite{Aspelmeyer2014RMP,fang2017generalized} and photonic lattices\cite{Porras2019,fitzpatrick2017observation}. The bosonic modes are denoted by $a_1,...,a_L$. For simplicity, let all modes have the same bare frequency $\omega_0$, and each mode is coupled to its neighbors with strengthes $t_1$ and $t_2$ [Fig. \ref{fig1}(d)]. Each site receives a coherent drive with amplitude $\epsilon_i(t)$, which can represent an incoming signal to be amplified. The Hamiltonian reads
\bea
H_{0}=&&\sum_{i}[(t_1a_i^\dagger a_{i+1}+t_2a_i^\dagger a_{i+2}+\text{H.c.})+\omega_0 a_i^\dag a_i \nn\\ && +\epsilon_i(t)a_i^\dagger+\epsilon_i^*(t)a_i].
\eea
As the system is open, we consider the density matrix $\rho$, whose time evolution follows the quantum master equation
\bea
\dot\rho(t)=-i[H_0,\rho]+\sum_\mu\left(L_\mu\rho L_\mu^\dagger-\frac{1}{2}\{L_\mu^\dagger L_\mu,\rho\}\right), \label{master}
\eea where $L_\mu$'s are the dissipators describing the effects of environment. While the physics is general, we take the following set of dissipators for concreteness: $\{L_\mu\}=\{L_i^{g},L_i^{l}\}$, including the single-particle gain $L_i^{g}=\sqrt{\gamma'}a_i^\dagger$ and loss $L_i^{l}=\sqrt\gamma( a_i-ia_{i+2})$. Feasible implementations of such dissipators have been discussed in details\cite{Metelmann2015,Porras2019}.

The most measurable quantity is the field coherence $\psi_i(t) =\langle a_i(t)\rangle = \text{Tr}[a_i\rho(t)]$. It follows from Eq. (\ref{master}) that they evolve under an effective non-Hermitian Hamiltonian $H$\cite{supplemental}:
\bea
\dot \psi_i=-i\sum_jH_{ij}\psi_j-i\epsilon_i. \label{Heff}
\eea
The same Eq. (\ref{Heff}) also generally arises in other physical platforms of directional amplification. Our results will be independent of specific implementation and applicable to a general $H$. For the specific model in Fig. \ref{fig1}(d), $H$ is found to be Fig. \ref{fig1}(a)\cite{supplemental}\footnote{It can be viewed as a boson counterpart of the fermion damping matrix\cite{Song2019}.}, with $\kappa =\frac{\gamma'}{2}-\gamma$.

Let us introduce the vector notation $\vec{\epsilon}=(\epsilon_1,\cdots,\epsilon_L)^T$, and similarly for $\vec{\psi}$.
For a signal $\vec{\epsilon}$ with a frequency $\omega$, $\vec{\epsilon}(t)=\vec{\epsilon}(\omega)\exp(-i\omega t)$, the resultant field is $\vec{\psi}(t)=\vec{\psi}(\omega)\exp(-i\omega t)$ whose amplitude is
\bea
\vec{\psi}(\omega)= G(\omega)\vec{\epsilon}(\omega); \quad G(\omega)=\frac{1}{\omega -H}.
\eea  To simplify notations, we shall measure the frequency with respect to $\omega_0$, namely, rename $\omega-\omega_0$ as $\omega$. Mathematically, this is equivalent to taking $\omega_0=0$. As such, negative $\omega$ denotes frequencies lower than $\omega_0$.

It is now evident that the Green's function matrix $G$ determines the amplification. For a signal entering the $j$ site, with the only nonzero component of $\vec{\epsilon}$ being $\epsilon_j$, the induced field at $i$ site is $\psi_i(\omega)=G_{ij}(\omega)\epsilon_j(\omega)$. Note that in the input-output formalism of amplification\cite{Clerk2010}, the scattering matrix $S$ is not exactly the same as, but has a simple relation to the Green's function: $S_{ij}(\omega)=\delta_{ij}-i\mu_i \mu_jG_{ij}(\omega)$, with certain $L$-independent coefficients $\mu_{i,j}$\cite{McDonald2018phase,Wanjura2019,ranzani2015graph}. Therefore, $S_{1L}$ ($S_{L1}$) is simply proportional to $G_{1L}$ ($G_{L1}$), and it suffices to focus on $G$.

{\it Discussions.--}We have obtained general formulas of Green's function for 1D non-Hermitian systems. As a practical application, our results serve as simple formulas of the gain and directionality of directional amplification, which provide an efficient guide for designing high-quality directional amplifiers. The general applicability of our formulas is independent of the specific physical platform. Moreover, in view of the versatile roles of Green's function in Hermitian systems, the general formula of Green's function obtained here is expected to have various applications in non-Hermitian bands. For example, it will be useful in studying the interaction effects that are significant in many open hybrid systems.

{\it Acknowledgements.--}We would like to thank Hongyi Wang for helpful discussions. This work is supported by NSFC under Grant No. 11674189.
WTX, MRL, YMH, and FS contributed equally to this work.

\bibliography{dirac}

\begin{thebibliography}{58}%
\makeatletter
\providecommand \@ifxundefined [1]{%
 \@ifx{#1\undefined}
}%
\providecommand \@ifnum [1]{%
 \ifnum #1\expandafter \@firstoftwo
 \else \expandafter \@secondoftwo
 \fi
}%
\providecommand \@ifx [1]{%
 \ifx #1\expandafter \@firstoftwo
 \else \expandafter \@secondoftwo
 \fi
}%
\providecommand \natexlab [1]{#1}%
\providecommand \enquote  [1]{``#1''}%
\providecommand \bibnamefont  [1]{#1}%
\providecommand \bibfnamefont [1]{#1}%
\providecommand \citenamefont [1]{#1}%
\providecommand \href@noop [0]{\@secondoftwo}%
\providecommand \href [0]{\begingroup \@sanitize@url \@href}%
\providecommand \@href[1]{\@@startlink{#1}\@@href}%
\providecommand \@@href[1]{\endgroup#1\@@endlink}%
\providecommand \@sanitize@url [0]{\catcode `\\12\catcode `\$12\catcode
  `\&12\catcode `\#12\catcode `\^12\catcode `\_12\catcode `\%12\relax}%
\providecommand \@@startlink[1]{}%
\providecommand \@@endlink[0]{}%
\providecommand \url  [0]{\begingroup\@sanitize@url \@url }%
\providecommand \@url [1]{\endgroup\@href {#1}{\urlprefix }}%
\providecommand \urlprefix  [0]{URL }%
\providecommand \Eprint [0]{\href }%
\providecommand \doibase [0]{http://dx.doi.org/}%
\providecommand \selectlanguage [0]{\@gobble}%
\providecommand \bibinfo  [0]{\@secondoftwo}%
\providecommand \bibfield  [0]{\@secondoftwo}%
\providecommand \translation [1]{[#1]}%
\providecommand \BibitemOpen [0]{}%
\providecommand \bibitemStop [0]{}%
\providecommand \bibitemNoStop [0]{.\EOS\space}%
\providecommand \EOS [0]{\spacefactor3000\relax}%
\providecommand \BibitemShut  [1]{\csname bibitem#1\endcsname}%
\let\auto@bib@innerbib\@empty
\bibitem [{\citenamefont {Yao}\ and\ \citenamefont {Wang}(2018)}]{yao2018edge}%
  \BibitemOpen
  \bibfield  {author} {\bibinfo {author} {\bibfnamefont {Shunyu}\ \bibnamefont
  {Yao}}\ and\ \bibinfo {author} {\bibfnamefont {Zhong}\ \bibnamefont {Wang}},\
  }\bibfield  {title} {\enquote {\bibinfo {title} {Edge states and topological
  invariants of non-hermitian systems},}\ }\href {\doibase
  10.1103/PhysRevLett.121.086803} {\bibfield  {journal} {\bibinfo  {journal}
  {Phys. Rev. Lett.}\ }\textbf {\bibinfo {volume} {121}},\ \bibinfo {pages}
  {086803} (\bibinfo {year} {2018})}\BibitemShut {NoStop}%
\bibitem [{\citenamefont {Kunst}\ \emph {et~al.}(2018)\citenamefont {Kunst},
  \citenamefont {Edvardsson}, \citenamefont {Budich},\ and\ \citenamefont
  {Bergholtz}}]{kunst2018biorthogonal}%
  \BibitemOpen
  \bibfield  {author} {\bibinfo {author} {\bibfnamefont {Flore~K.}\
  \bibnamefont {Kunst}}, \bibinfo {author} {\bibfnamefont {Elisabet}\
  \bibnamefont {Edvardsson}}, \bibinfo {author} {\bibfnamefont {Jan~Carl}\
  \bibnamefont {Budich}}, \ and\ \bibinfo {author} {\bibfnamefont {Emil~J.}\
  \bibnamefont {Bergholtz}},\ }\bibfield  {title} {\enquote {\bibinfo {title}
  {Biorthogonal bulk-boundary correspondence in non-hermitian systems},}\
  }\href {\doibase 10.1103/PhysRevLett.121.026808} {\bibfield  {journal}
  {\bibinfo  {journal} {Phys. Rev. Lett.}\ }\textbf {\bibinfo {volume} {121}},\
  \bibinfo {pages} {026808} (\bibinfo {year} {2018})}\BibitemShut {NoStop}%
\bibitem [{\citenamefont {Lee}\ and\ \citenamefont
  {Thomale}(2019)}]{lee2018anatomy}%
  \BibitemOpen
  \bibfield  {author} {\bibinfo {author} {\bibfnamefont {Ching~Hua}\
  \bibnamefont {Lee}}\ and\ \bibinfo {author} {\bibfnamefont {Ronny}\
  \bibnamefont {Thomale}},\ }\bibfield  {title} {\enquote {\bibinfo {title}
  {Anatomy of skin modes and topology in non-hermitian systems},}\ }\href
  {\doibase 10.1103/PhysRevB.99.201103} {\bibfield  {journal} {\bibinfo
  {journal} {Phys. Rev. B}\ }\textbf {\bibinfo {volume} {99}},\ \bibinfo
  {pages} {201103} (\bibinfo {year} {2019})}\BibitemShut {NoStop}%
\bibitem [{\citenamefont {Martinez~Alvarez}\ \emph {et~al.}(2018)\citenamefont
  {Martinez~Alvarez}, \citenamefont {Barrios~Vargas},\ and\ \citenamefont
  {Foa~Torres}}]{alvarez2017}%
  \BibitemOpen
  \bibfield  {author} {\bibinfo {author} {\bibfnamefont {V.~M.}\ \bibnamefont
  {Martinez~Alvarez}}, \bibinfo {author} {\bibfnamefont {J.~E.}\ \bibnamefont
  {Barrios~Vargas}}, \ and\ \bibinfo {author} {\bibfnamefont {L.~E.~F.}\
  \bibnamefont {Foa~Torres}},\ }\bibfield  {title} {\enquote {\bibinfo {title}
  {Non-hermitian robust edge states in one dimension: Anomalous localization
  and eigenspace condensation at exceptional points},}\ }\href {\doibase
  10.1103/PhysRevB.97.121401} {\bibfield  {journal} {\bibinfo  {journal} {Phys.
  Rev. B}\ }\textbf {\bibinfo {volume} {97}},\ \bibinfo {pages} {121401}
  (\bibinfo {year} {2018})}\BibitemShut {NoStop}%
\bibitem [{\citenamefont {{Xiao}}\ \emph {et~al.}(2020)\citenamefont {{Xiao}},
  \citenamefont {{Deng}}, \citenamefont {{Wang}}, \citenamefont {{Zhu}},
  \citenamefont {{Wang}}, \citenamefont {{Yi}},\ and\ \citenamefont
  {{Xue}}}]{Xiao2019NHSE}%
  \BibitemOpen
  \bibfield  {author} {\bibinfo {author} {\bibfnamefont {Lei}\ \bibnamefont
  {{Xiao}}}, \bibinfo {author} {\bibfnamefont {Tianshu}\ \bibnamefont
  {{Deng}}}, \bibinfo {author} {\bibfnamefont {Kunkun}\ \bibnamefont {{Wang}}},
  \bibinfo {author} {\bibfnamefont {Gaoyan}\ \bibnamefont {{Zhu}}}, \bibinfo
  {author} {\bibfnamefont {Zhong}\ \bibnamefont {{Wang}}}, \bibinfo {author}
  {\bibfnamefont {Wei}\ \bibnamefont {{Yi}}}, \ and\ \bibinfo {author}
  {\bibfnamefont {Peng}\ \bibnamefont {{Xue}}},\ }\bibfield  {title} {\enquote
  {\bibinfo {title} {{Non-Hermitian bulk-boundary correspondence in quantum
  dynamics}},}\ }\href@noop {} {\bibfield  {journal} {\bibinfo  {journal}
  {Nature Physics}\ }\textbf {\bibinfo {volume} {16}},\ \bibinfo {pages} {761}
  (\bibinfo {year} {2020})},\ \Eprint {http://arxiv.org/abs/1907.12566}
  {1907.12566 [cond-mat.mes-hall]} \BibitemShut {NoStop}%
\bibitem [{\citenamefont {Helbig}\ \emph {et~al.}(2020)\citenamefont {Helbig},
  \citenamefont {Hofmann}, \citenamefont {Imhof}, \citenamefont {Abdelghany},
  \citenamefont {Kiessling}, \citenamefont {Molenkamp}, \citenamefont {Lee},
  \citenamefont {Szameit}, \citenamefont {Greiter},\ and\ \citenamefont
  {Thomale}}]{Helbig2019NHSE}%
  \BibitemOpen
  \bibfield  {author} {\bibinfo {author} {\bibfnamefont {T.}~\bibnamefont
  {Helbig}}, \bibinfo {author} {\bibfnamefont {T.}~\bibnamefont {Hofmann}},
  \bibinfo {author} {\bibfnamefont {S.}~\bibnamefont {Imhof}}, \bibinfo
  {author} {\bibfnamefont {M.}~\bibnamefont {Abdelghany}}, \bibinfo {author}
  {\bibfnamefont {T.}~\bibnamefont {Kiessling}}, \bibinfo {author}
  {\bibfnamefont {L.~W.}\ \bibnamefont {Molenkamp}}, \bibinfo {author}
  {\bibfnamefont {C.~H.}\ \bibnamefont {Lee}}, \bibinfo {author} {\bibfnamefont
  {A.}~\bibnamefont {Szameit}}, \bibinfo {author} {\bibfnamefont
  {M.}~\bibnamefont {Greiter}}, \ and\ \bibinfo {author} {\bibfnamefont
  {R.}~\bibnamefont {Thomale}},\ }\bibfield  {title} {\enquote {\bibinfo
  {title} {Generalized bulk--boundary correspondence in non-hermitian
  topolectrical circuits},}\ }\href@noop {} {\bibfield  {journal} {\bibinfo
  {journal} {Nature Physics}\ }\textbf {\bibinfo {volume} {16}},\ \bibinfo
  {pages} {747} (\bibinfo {year} {2020})}\BibitemShut {NoStop}%
\bibitem [{\citenamefont {Ghatak}\ \emph {et~al.}(2020)\citenamefont {Ghatak},
  \citenamefont {Brandenbourger}, \citenamefont {van Wezel},\ and\
  \citenamefont {Coulais}}]{Ghatak2019NHSE}%
  \BibitemOpen
  \bibfield  {author} {\bibinfo {author} {\bibfnamefont {Ananya}\ \bibnamefont
  {Ghatak}}, \bibinfo {author} {\bibfnamefont {Martin}\ \bibnamefont
  {Brandenbourger}}, \bibinfo {author} {\bibfnamefont {Jasper}\ \bibnamefont
  {van Wezel}}, \ and\ \bibinfo {author} {\bibfnamefont {Corentin}\
  \bibnamefont {Coulais}},\ }\bibfield  {title} {\enquote {\bibinfo {title}
  {Observation of non-hermitian topology and its bulk--edge correspondence in
  an active mechanical metamaterial},}\ }\href@noop {} {\bibfield  {journal}
  {\bibinfo  {journal} {Proceedings of the National Academy of Sciences}\
  }\textbf {\bibinfo {volume} {117}},\ \bibinfo {pages} {29561--29568}
  (\bibinfo {year} {2020})}\BibitemShut {NoStop}%
\bibitem [{Note1()}]{Note1}%
  \BibitemOpen
  \bibinfo {note} {Our results will be generalizable to other boundary
  conditions such as domain wall systems. In mathematical literature, OBC
  corresponds to Toeplitz matrices, while PBC corresponds to circulant
  matrices.}\BibitemShut {Stop}%
\bibitem [{\citenamefont {Borgnia}\ \emph {et~al.}(2020)\citenamefont
  {Borgnia}, \citenamefont {Kruchkov},\ and\ \citenamefont
  {Slager}}]{Borgnia2019}%
  \BibitemOpen
  \bibfield  {author} {\bibinfo {author} {\bibfnamefont {Dan~S.}\ \bibnamefont
  {Borgnia}}, \bibinfo {author} {\bibfnamefont {Alex~Jura}\ \bibnamefont
  {Kruchkov}}, \ and\ \bibinfo {author} {\bibfnamefont {Robert-Jan}\
  \bibnamefont {Slager}},\ }\bibfield  {title} {\enquote {\bibinfo {title}
  {Non-hermitian boundary modes and topology},}\ }\href {\doibase
  10.1103/PhysRevLett.124.056802} {\bibfield  {journal} {\bibinfo  {journal}
  {Phys. Rev. Lett.}\ }\textbf {\bibinfo {volume} {124}},\ \bibinfo {pages}
  {056802} (\bibinfo {year} {2020})}\BibitemShut {NoStop}%
\bibitem [{\citenamefont {{Zirnstein}}\ \emph {et~al.}(2019)\citenamefont
  {{Zirnstein}}, \citenamefont {{Refael}},\ and\ \citenamefont
  {{Rosenow}}}]{Zirnstein2019}%
  \BibitemOpen
  \bibfield  {author} {\bibinfo {author} {\bibfnamefont {Heinrich-Gregor}\
  \bibnamefont {{Zirnstein}}}, \bibinfo {author} {\bibfnamefont {Gil}\
  \bibnamefont {{Refael}}}, \ and\ \bibinfo {author} {\bibfnamefont {Bernd}\
  \bibnamefont {{Rosenow}}},\ }\bibfield  {title} {\enquote {\bibinfo {title}
  {{Bulk-boundary correspondence for non-Hermitian Hamiltonians via Green
  functions}},}\ }\href@noop {} {\bibfield  {journal} {\bibinfo  {journal}
  {arXiv e-prints}\ ,\ \bibinfo {eid} {arXiv:1901.11241}} (\bibinfo {year}
  {2019})},\ \Eprint {http://arxiv.org/abs/1901.11241} {arXiv:1901.11241
  [cond-mat.mes-hall]} \BibitemShut {NoStop}%
\bibitem [{\citenamefont {Silveirinha}(2019)}]{Silveirinha2019}%
  \BibitemOpen
  \bibfield  {author} {\bibinfo {author} {\bibfnamefont {M\'ario~G.}\
  \bibnamefont {Silveirinha}},\ }\bibfield  {title} {\enquote {\bibinfo {title}
  {Topological theory of non-hermitian photonic systems},}\ }\href {\doibase
  10.1103/PhysRevB.99.125155} {\bibfield  {journal} {\bibinfo  {journal} {Phys.
  Rev. B}\ }\textbf {\bibinfo {volume} {99}},\ \bibinfo {pages} {125155}
  (\bibinfo {year} {2019})}\BibitemShut {NoStop}%
\bibitem [{\citenamefont {Yokomizo}\ and\ \citenamefont
  {Murakami}(2019)}]{Yokomizo2019}%
  \BibitemOpen
  \bibfield  {author} {\bibinfo {author} {\bibfnamefont {Kazuki}\ \bibnamefont
  {Yokomizo}}\ and\ \bibinfo {author} {\bibfnamefont {Shuichi}\ \bibnamefont
  {Murakami}},\ }\bibfield  {title} {\enquote {\bibinfo {title} {Non-bloch band
  theory of non-hermitian systems},}\ }\href {\doibase
  10.1103/PhysRevLett.123.066404} {\bibfield  {journal} {\bibinfo  {journal}
  {Phys. Rev. Lett.}\ }\textbf {\bibinfo {volume} {123}},\ \bibinfo {pages}
  {066404} (\bibinfo {year} {2019})}\BibitemShut {NoStop}%
\bibitem [{\citenamefont {Yao}\ \emph {et~al.}(2018)\citenamefont {Yao},
  \citenamefont {Song},\ and\ \citenamefont {Wang}}]{yao2018chern}%
  \BibitemOpen
  \bibfield  {author} {\bibinfo {author} {\bibfnamefont {Shunyu}\ \bibnamefont
  {Yao}}, \bibinfo {author} {\bibfnamefont {Fei}\ \bibnamefont {Song}}, \ and\
  \bibinfo {author} {\bibfnamefont {Zhong}\ \bibnamefont {Wang}},\ }\bibfield
  {title} {\enquote {\bibinfo {title} {Non-hermitian chern bands},}\ }\href
  {\doibase 10.1103/PhysRevLett.121.136802} {\bibfield  {journal} {\bibinfo
  {journal} {Phys. Rev. Lett.}\ }\textbf {\bibinfo {volume} {121}},\ \bibinfo
  {pages} {136802} (\bibinfo {year} {2018})}\BibitemShut {NoStop}%
\bibitem [{\citenamefont {Longhi}(2019)}]{Longhi2019Probing}%
  \BibitemOpen
  \bibfield  {author} {\bibinfo {author} {\bibfnamefont {Stefano}\ \bibnamefont
  {Longhi}},\ }\bibfield  {title} {\enquote {\bibinfo {title} {Probing
  non-hermitian skin effect and non-bloch phase transitions},}\ }\href
  {\doibase 10.1103/PhysRevResearch.1.023013} {\bibfield  {journal} {\bibinfo
  {journal} {Phys. Rev. Research}\ }\textbf {\bibinfo {volume} {1}},\ \bibinfo
  {pages} {023013} (\bibinfo {year} {2019})}\BibitemShut {NoStop}%
\bibitem [{\citenamefont {Yang}\ \emph {et~al.}(2020)\citenamefont {Yang},
  \citenamefont {Zhang}, \citenamefont {Fang},\ and\ \citenamefont
  {Hu}}]{Yang2019Auxiliary}%
  \BibitemOpen
  \bibfield  {author} {\bibinfo {author} {\bibfnamefont {Zhesen}\ \bibnamefont
  {Yang}}, \bibinfo {author} {\bibfnamefont {Kai}\ \bibnamefont {Zhang}},
  \bibinfo {author} {\bibfnamefont {Chen}\ \bibnamefont {Fang}}, \ and\
  \bibinfo {author} {\bibfnamefont {Jiangping}\ \bibnamefont {Hu}},\ }\bibfield
   {title} {\enquote {\bibinfo {title} {Non-hermitian bulk-boundary
  correspondence and auxiliary generalized brillouin zone theory},}\ }\href
  {\doibase 10.1103/PhysRevLett.125.226402} {\bibfield  {journal} {\bibinfo
  {journal} {Phys. Rev. Lett.}\ }\textbf {\bibinfo {volume} {125}},\ \bibinfo
  {pages} {226402} (\bibinfo {year} {2020})}\BibitemShut {NoStop}%
\bibitem [{\citenamefont {Deng}\ and\ \citenamefont {Yi}(2019)}]{Deng2019}%
  \BibitemOpen
  \bibfield  {author} {\bibinfo {author} {\bibfnamefont {Tian-Shu}\
  \bibnamefont {Deng}}\ and\ \bibinfo {author} {\bibfnamefont {Wei}\
  \bibnamefont {Yi}},\ }\bibfield  {title} {\enquote {\bibinfo {title}
  {Non-bloch topological invariants in a non-hermitian domain wall system},}\
  }\href {\doibase 10.1103/PhysRevB.100.035102} {\bibfield  {journal} {\bibinfo
   {journal} {Phys. Rev. B}\ }\textbf {\bibinfo {volume} {100}},\ \bibinfo
  {pages} {035102} (\bibinfo {year} {2019})}\BibitemShut {NoStop}%
\bibitem [{\citenamefont {Longhi}(2020)}]{Longhi2020chiral}%
  \BibitemOpen
  \bibfield  {author} {\bibinfo {author} {\bibfnamefont {S.}~\bibnamefont
  {Longhi}},\ }\bibfield  {title} {\enquote {\bibinfo {title} {Non-bloch-band
  collapse and chiral zener tunneling},}\ }\href {\doibase
  10.1103/PhysRevLett.124.066602} {\bibfield  {journal} {\bibinfo  {journal}
  {Phys. Rev. Lett.}\ }\textbf {\bibinfo {volume} {124}},\ \bibinfo {pages}
  {066602} (\bibinfo {year} {2020})}\BibitemShut {NoStop}%
\bibitem [{\citenamefont {Kawabata}\ \emph {et~al.}(2020)\citenamefont
  {Kawabata}, \citenamefont {Okuma},\ and\ \citenamefont
  {Sato}}]{Kawabata2020nonBloch}%
  \BibitemOpen
  \bibfield  {author} {\bibinfo {author} {\bibfnamefont {Kohei}\ \bibnamefont
  {Kawabata}}, \bibinfo {author} {\bibfnamefont {Nobuyuki}\ \bibnamefont
  {Okuma}}, \ and\ \bibinfo {author} {\bibfnamefont {Masatoshi}\ \bibnamefont
  {Sato}},\ }\bibfield  {title} {\enquote {\bibinfo {title} {Non-bloch band
  theory of non-hermitian hamiltonians in the symplectic class},}\ }\href
  {\doibase 10.1103/PhysRevB.101.195147} {\bibfield  {journal} {\bibinfo
  {journal} {Phys. Rev. B}\ }\textbf {\bibinfo {volume} {101}},\ \bibinfo
  {pages} {195147} (\bibinfo {year} {2020})}\BibitemShut {NoStop}%
\bibitem [{\citenamefont {Liu}\ \emph {et~al.}(2019)\citenamefont {Liu},
  \citenamefont {Zhang}, \citenamefont {Ai}, \citenamefont {Gong},
  \citenamefont {Kawabata}, \citenamefont {Ueda},\ and\ \citenamefont
  {Nori}}]{liu2019second}%
  \BibitemOpen
  \bibfield  {author} {\bibinfo {author} {\bibfnamefont {Tao}\ \bibnamefont
  {Liu}}, \bibinfo {author} {\bibfnamefont {Yu-Ran}\ \bibnamefont {Zhang}},
  \bibinfo {author} {\bibfnamefont {Qing}\ \bibnamefont {Ai}}, \bibinfo
  {author} {\bibfnamefont {Zongping}\ \bibnamefont {Gong}}, \bibinfo {author}
  {\bibfnamefont {Kohei}\ \bibnamefont {Kawabata}}, \bibinfo {author}
  {\bibfnamefont {Masahito}\ \bibnamefont {Ueda}}, \ and\ \bibinfo {author}
  {\bibfnamefont {Franco}\ \bibnamefont {Nori}},\ }\bibfield  {title} {\enquote
  {\bibinfo {title} {Second-order topological phases in non-hermitian
  systems},}\ }\href {\doibase 10.1103/PhysRevLett.122.076801} {\bibfield
  {journal} {\bibinfo  {journal} {Phys. Rev. Lett.}\ }\textbf {\bibinfo
  {volume} {122}},\ \bibinfo {pages} {076801} (\bibinfo {year}
  {2019})}\BibitemShut {NoStop}%
\bibitem [{\citenamefont {Song}\ \emph
  {et~al.}(2019{\natexlab{a}})\citenamefont {Song}, \citenamefont {Yao},\ and\
  \citenamefont {Wang}}]{Song2019real}%
  \BibitemOpen
  \bibfield  {author} {\bibinfo {author} {\bibfnamefont {Fei}\ \bibnamefont
  {Song}}, \bibinfo {author} {\bibfnamefont {Shunyu}\ \bibnamefont {Yao}}, \
  and\ \bibinfo {author} {\bibfnamefont {Zhong}\ \bibnamefont {Wang}},\
  }\bibfield  {title} {\enquote {\bibinfo {title} {Non-hermitian topological
  invariants in real space},}\ }\href {\doibase 10.1103/PhysRevLett.123.246801}
  {\bibfield  {journal} {\bibinfo  {journal} {Phys. Rev. Lett.}\ }\textbf
  {\bibinfo {volume} {123}},\ \bibinfo {pages} {246801} (\bibinfo {year}
  {2019}{\natexlab{a}})}\BibitemShut {NoStop}%
\bibitem [{\citenamefont {Lee}\ \emph {et~al.}(2020)\citenamefont {Lee},
  \citenamefont {Li}, \citenamefont {Thomale},\ and\ \citenamefont
  {Gong}}]{Lee2020Unraveling}%
  \BibitemOpen
  \bibfield  {author} {\bibinfo {author} {\bibfnamefont {Ching~Hua}\
  \bibnamefont {Lee}}, \bibinfo {author} {\bibfnamefont {Linhu}\ \bibnamefont
  {Li}}, \bibinfo {author} {\bibfnamefont {Ronny}\ \bibnamefont {Thomale}}, \
  and\ \bibinfo {author} {\bibfnamefont {Jiangbin}\ \bibnamefont {Gong}},\
  }\bibfield  {title} {\enquote {\bibinfo {title} {Unraveling non-hermitian
  pumping: Emergent spectral singularities and anomalous responses},}\ }\href
  {\doibase 10.1103/PhysRevB.102.085151} {\bibfield  {journal} {\bibinfo
  {journal} {Phys. Rev. B}\ }\textbf {\bibinfo {volume} {102}},\ \bibinfo
  {pages} {085151} (\bibinfo {year} {2020})}\BibitemShut {NoStop}%
\bibitem [{\citenamefont {Yi}\ and\ \citenamefont {Yang}(2020)}]{Yi2020}%
  \BibitemOpen
  \bibfield  {author} {\bibinfo {author} {\bibfnamefont {Yifei}\ \bibnamefont
  {Yi}}\ and\ \bibinfo {author} {\bibfnamefont {Zhesen}\ \bibnamefont {Yang}},\
  }\bibfield  {title} {\enquote {\bibinfo {title} {Non-hermitian skin modes
  induced by on-site dissipations and chiral tunneling effect},}\ }\href
  {\doibase 10.1103/PhysRevLett.125.186802} {\bibfield  {journal} {\bibinfo
  {journal} {Phys. Rev. Lett.}\ }\textbf {\bibinfo {volume} {125}},\ \bibinfo
  {pages} {186802} (\bibinfo {year} {2020})}\BibitemShut {NoStop}%
\bibitem [{\citenamefont {Bergholtz}\ \emph {et~al.}(2021)\citenamefont
  {Bergholtz}, \citenamefont {Budich},\ and\ \citenamefont
  {Kunst}}]{Bergholtz2020RMP}%
  \BibitemOpen
  \bibfield  {author} {\bibinfo {author} {\bibfnamefont {Emil~J.}\ \bibnamefont
  {Bergholtz}}, \bibinfo {author} {\bibfnamefont {Jan~Carl}\ \bibnamefont
  {Budich}}, \ and\ \bibinfo {author} {\bibfnamefont {Flore~K.}\ \bibnamefont
  {Kunst}},\ }\bibfield  {title} {\enquote {\bibinfo {title} {Exceptional
  topology of non-hermitian systems},}\ }\href {\doibase
  10.1103/RevModPhys.93.015005} {\bibfield  {journal} {\bibinfo  {journal}
  {Rev. Mod. Phys.}\ }\textbf {\bibinfo {volume} {93}},\ \bibinfo {pages}
  {015005} (\bibinfo {year} {2021})}\BibitemShut {NoStop}%
\bibitem [{\citenamefont {{Ashida}}\ \emph {et~al.}(2020)\citenamefont
  {{Ashida}}, \citenamefont {{Gong}},\ and\ \citenamefont
  {{Ueda}}}]{Ashida2020}%
  \BibitemOpen
  \bibfield  {author} {\bibinfo {author} {\bibfnamefont {Y.}~\bibnamefont
  {{Ashida}}}, \bibinfo {author} {\bibfnamefont {Z.}~\bibnamefont {{Gong}}}, \
  and\ \bibinfo {author} {\bibfnamefont {M.}~\bibnamefont {{Ueda}}},\
  }\bibfield  {title} {\enquote {\bibinfo {title} {{Non-Hermitian Physics}},}\
  }\href@noop {} {\bibfield  {journal} {\bibinfo  {journal} {arXiv e-prints}\
  ,\ \bibinfo {eid} {arXiv:2006.01837}} (\bibinfo {year} {2020})},\ \Eprint
  {http://arxiv.org/abs/2006.01837} {arXiv:2006.01837 [cond-mat.mes-hall]}
  \BibitemShut {NoStop}%
\bibitem [{\citenamefont {Abdo}\ \emph {et~al.}(2013)\citenamefont {Abdo},
  \citenamefont {Sliwa}, \citenamefont {Frunzio},\ and\ \citenamefont
  {Devoret}}]{Abdo2013directional}%
  \BibitemOpen
  \bibfield  {author} {\bibinfo {author} {\bibfnamefont {Baleegh}\ \bibnamefont
  {Abdo}}, \bibinfo {author} {\bibfnamefont {Katrina}\ \bibnamefont {Sliwa}},
  \bibinfo {author} {\bibfnamefont {Luigi}\ \bibnamefont {Frunzio}}, \ and\
  \bibinfo {author} {\bibfnamefont {Michel}\ \bibnamefont {Devoret}},\
  }\bibfield  {title} {\enquote {\bibinfo {title} {Directional amplification
  with a josephson circuit},}\ }\href {\doibase 10.1103/PhysRevX.3.031001}
  {\bibfield  {journal} {\bibinfo  {journal} {Phys. Rev. X}\ }\textbf {\bibinfo
  {volume} {3}},\ \bibinfo {pages} {031001} (\bibinfo {year}
  {2013})}\BibitemShut {NoStop}%
\bibitem [{\citenamefont {Abdo}\ \emph {et~al.}(2014)\citenamefont {Abdo},
  \citenamefont {Sliwa}, \citenamefont {Shankar}, \citenamefont {Hatridge},
  \citenamefont {Frunzio}, \citenamefont {Schoelkopf},\ and\ \citenamefont
  {Devoret}}]{Abdo2014}%
  \BibitemOpen
  \bibfield  {author} {\bibinfo {author} {\bibfnamefont {Baleegh}\ \bibnamefont
  {Abdo}}, \bibinfo {author} {\bibfnamefont {Katrina}\ \bibnamefont {Sliwa}},
  \bibinfo {author} {\bibfnamefont {S.}~\bibnamefont {Shankar}}, \bibinfo
  {author} {\bibfnamefont {Michael}\ \bibnamefont {Hatridge}}, \bibinfo
  {author} {\bibfnamefont {Luigi}\ \bibnamefont {Frunzio}}, \bibinfo {author}
  {\bibfnamefont {Robert}\ \bibnamefont {Schoelkopf}}, \ and\ \bibinfo {author}
  {\bibfnamefont {Michel}\ \bibnamefont {Devoret}},\ }\bibfield  {title}
  {\enquote {\bibinfo {title} {Josephson directional amplifier for quantum
  measurement of superconducting circuits},}\ }\href {\doibase
  10.1103/PhysRevLett.112.167701} {\bibfield  {journal} {\bibinfo  {journal}
  {Phys. Rev. Lett.}\ }\textbf {\bibinfo {volume} {112}},\ \bibinfo {pages}
  {167701} (\bibinfo {year} {2014})}\BibitemShut {NoStop}%
\bibitem [{\citenamefont {Sliwa}\ \emph {et~al.}(2015)\citenamefont {Sliwa},
  \citenamefont {Hatridge}, \citenamefont {Narla}, \citenamefont {Shankar},
  \citenamefont {Frunzio}, \citenamefont {Schoelkopf},\ and\ \citenamefont
  {Devoret}}]{Sliwa2015Reconfigurable}%
  \BibitemOpen
  \bibfield  {author} {\bibinfo {author} {\bibfnamefont {K.~M.}\ \bibnamefont
  {Sliwa}}, \bibinfo {author} {\bibfnamefont {M.}~\bibnamefont {Hatridge}},
  \bibinfo {author} {\bibfnamefont {A.}~\bibnamefont {Narla}}, \bibinfo
  {author} {\bibfnamefont {S.}~\bibnamefont {Shankar}}, \bibinfo {author}
  {\bibfnamefont {L.}~\bibnamefont {Frunzio}}, \bibinfo {author} {\bibfnamefont
  {R.~J.}\ \bibnamefont {Schoelkopf}}, \ and\ \bibinfo {author} {\bibfnamefont
  {M.~H.}\ \bibnamefont {Devoret}},\ }\bibfield  {title} {\enquote {\bibinfo
  {title} {Reconfigurable josephson circulator/directional amplifier},}\ }\href
  {\doibase 10.1103/PhysRevX.5.041020} {\bibfield  {journal} {\bibinfo
  {journal} {Phys. Rev. X}\ }\textbf {\bibinfo {volume} {5}},\ \bibinfo {pages}
  {041020} (\bibinfo {year} {2015})}\BibitemShut {NoStop}%
\bibitem [{\citenamefont {Jalas}\ \emph {et~al.}(2013)\citenamefont {Jalas},
  \citenamefont {Petrov}, \citenamefont {Eich}, \citenamefont {Freude},
  \citenamefont {Fan}, \citenamefont {Yu}, \citenamefont {Baets}, \citenamefont
  {Popovi{\'c}}, \citenamefont {Melloni}, \citenamefont {Joannopoulos} \emph
  {et~al.}}]{jalas2013what}%
  \BibitemOpen
  \bibfield  {author} {\bibinfo {author} {\bibfnamefont {Dirk}\ \bibnamefont
  {Jalas}}, \bibinfo {author} {\bibfnamefont {Alexander}\ \bibnamefont
  {Petrov}}, \bibinfo {author} {\bibfnamefont {Manfred}\ \bibnamefont {Eich}},
  \bibinfo {author} {\bibfnamefont {Wolfgang}\ \bibnamefont {Freude}}, \bibinfo
  {author} {\bibfnamefont {Shanhui}\ \bibnamefont {Fan}}, \bibinfo {author}
  {\bibfnamefont {Zongfu}\ \bibnamefont {Yu}}, \bibinfo {author} {\bibfnamefont
  {Roel}\ \bibnamefont {Baets}}, \bibinfo {author} {\bibfnamefont
  {Milo{\v{s}}}\ \bibnamefont {Popovi{\'c}}}, \bibinfo {author} {\bibfnamefont
  {Andrea}\ \bibnamefont {Melloni}}, \bibinfo {author} {\bibfnamefont {John~D}\
  \bibnamefont {Joannopoulos}},  \emph {et~al.},\ }\bibfield  {title} {\enquote
  {\bibinfo {title} {What is-and what is not-an optical isolator},}\
  }\href@noop {} {\bibfield  {journal} {\bibinfo  {journal} {Nature Photonics}\
  }\textbf {\bibinfo {volume} {7}},\ \bibinfo {pages} {579--582} (\bibinfo
  {year} {2013})}\BibitemShut {NoStop}%
\bibitem [{\citenamefont {Feng}\ \emph {et~al.}(2011)\citenamefont {Feng},
  \citenamefont {Ayache}, \citenamefont {Huang}, \citenamefont {Xu},
  \citenamefont {Lu}, \citenamefont {Chen}, \citenamefont {Fainman},\ and\
  \citenamefont {Scherer}}]{feng2011nonreciprocal}%
  \BibitemOpen
  \bibfield  {author} {\bibinfo {author} {\bibfnamefont {Liang}\ \bibnamefont
  {Feng}}, \bibinfo {author} {\bibfnamefont {Maurice}\ \bibnamefont {Ayache}},
  \bibinfo {author} {\bibfnamefont {Jingqing}\ \bibnamefont {Huang}}, \bibinfo
  {author} {\bibfnamefont {Ye-Long}\ \bibnamefont {Xu}}, \bibinfo {author}
  {\bibfnamefont {Ming-Hui}\ \bibnamefont {Lu}}, \bibinfo {author}
  {\bibfnamefont {Yan-Feng}\ \bibnamefont {Chen}}, \bibinfo {author}
  {\bibfnamefont {Yeshaiahu}\ \bibnamefont {Fainman}}, \ and\ \bibinfo {author}
  {\bibfnamefont {Axel}\ \bibnamefont {Scherer}},\ }\bibfield  {title}
  {\enquote {\bibinfo {title} {Nonreciprocal light propagation in a silicon
  photonic circuit},}\ }\href@noop {} {\bibfield  {journal} {\bibinfo
  {journal} {Science}\ }\textbf {\bibinfo {volume} {333}},\ \bibinfo {pages}
  {729--733} (\bibinfo {year} {2011})}\BibitemShut {NoStop}%
\bibitem [{\citenamefont {Caloz}\ \emph {et~al.}(2018)\citenamefont {Caloz},
  \citenamefont {Al\`u}, \citenamefont {Tretyakov}, \citenamefont {Sounas},
  \citenamefont {Achouri},\ and\ \citenamefont {Deck-L\'eger}}]{Caloz2018}%
  \BibitemOpen
  \bibfield  {author} {\bibinfo {author} {\bibfnamefont {Christophe}\
  \bibnamefont {Caloz}}, \bibinfo {author} {\bibfnamefont {Andrea}\
  \bibnamefont {Al\`u}}, \bibinfo {author} {\bibfnamefont {Sergei}\
  \bibnamefont {Tretyakov}}, \bibinfo {author} {\bibfnamefont {Dimitrios}\
  \bibnamefont {Sounas}}, \bibinfo {author} {\bibfnamefont {Karim}\
  \bibnamefont {Achouri}}, \ and\ \bibinfo {author} {\bibfnamefont
  {Zo\'e-Lise}\ \bibnamefont {Deck-L\'eger}},\ }\bibfield  {title} {\enquote
  {\bibinfo {title} {Electromagnetic nonreciprocity},}\ }\href {\doibase
  10.1103/PhysRevApplied.10.047001} {\bibfield  {journal} {\bibinfo  {journal}
  {Phys. Rev. Applied}\ }\textbf {\bibinfo {volume} {10}},\ \bibinfo {pages}
  {047001} (\bibinfo {year} {2018})}\BibitemShut {NoStop}%
\bibitem [{\citenamefont {Fleury}\ \emph {et~al.}(2014)\citenamefont {Fleury},
  \citenamefont {Sounas}, \citenamefont {Sieck}, \citenamefont {Haberman},\
  and\ \citenamefont {Al{\`u}}}]{fleury2014sound}%
  \BibitemOpen
  \bibfield  {author} {\bibinfo {author} {\bibfnamefont {Romain}\ \bibnamefont
  {Fleury}}, \bibinfo {author} {\bibfnamefont {Dimitrios~L}\ \bibnamefont
  {Sounas}}, \bibinfo {author} {\bibfnamefont {Caleb~F}\ \bibnamefont {Sieck}},
  \bibinfo {author} {\bibfnamefont {Michael~R}\ \bibnamefont {Haberman}}, \
  and\ \bibinfo {author} {\bibfnamefont {Andrea}\ \bibnamefont {Al{\`u}}},\
  }\bibfield  {title} {\enquote {\bibinfo {title} {Sound isolation and giant
  linear nonreciprocity in a compact acoustic circulator},}\ }\href@noop {}
  {\bibfield  {journal} {\bibinfo  {journal} {Science}\ }\textbf {\bibinfo
  {volume} {343}},\ \bibinfo {pages} {516--519} (\bibinfo {year}
  {2014})}\BibitemShut {NoStop}%
\bibitem [{\citenamefont {Sounas}\ and\ \citenamefont
  {Al{\`u}}(2017)}]{sounas2017non}%
  \BibitemOpen
  \bibfield  {author} {\bibinfo {author} {\bibfnamefont {Dimitrios~L}\
  \bibnamefont {Sounas}}\ and\ \bibinfo {author} {\bibfnamefont {Andrea}\
  \bibnamefont {Al{\`u}}},\ }\bibfield  {title} {\enquote {\bibinfo {title}
  {Non-reciprocal photonics based on time modulation},}\ }\href@noop {}
  {\bibfield  {journal} {\bibinfo  {journal} {Nature Photonics}\ }\textbf
  {\bibinfo {volume} {11}},\ \bibinfo {pages} {774--783} (\bibinfo {year}
  {2017})}\BibitemShut {NoStop}%
\bibitem [{\citenamefont {Yu}\ and\ \citenamefont
  {Fan}(2009)}]{yu2009complete}%
  \BibitemOpen
  \bibfield  {author} {\bibinfo {author} {\bibfnamefont {Zongfu}\ \bibnamefont
  {Yu}}\ and\ \bibinfo {author} {\bibfnamefont {Shanhui}\ \bibnamefont {Fan}},\
  }\bibfield  {title} {\enquote {\bibinfo {title} {Complete optical isolation
  created by indirect interband photonic transitions},}\ }\href@noop {}
  {\bibfield  {journal} {\bibinfo  {journal} {Nature photonics}\ }\textbf
  {\bibinfo {volume} {3}},\ \bibinfo {pages} {91} (\bibinfo {year}
  {2009})}\BibitemShut {NoStop}%
\bibitem [{\citenamefont {Estep}\ \emph {et~al.}(2014)\citenamefont {Estep},
  \citenamefont {Sounas}, \citenamefont {Soric},\ and\ \citenamefont
  {Al{\`u}}}]{estep2014magnetic}%
  \BibitemOpen
  \bibfield  {author} {\bibinfo {author} {\bibfnamefont {Nicholas~A}\
  \bibnamefont {Estep}}, \bibinfo {author} {\bibfnamefont {Dimitrios~L}\
  \bibnamefont {Sounas}}, \bibinfo {author} {\bibfnamefont {Jason}\
  \bibnamefont {Soric}}, \ and\ \bibinfo {author} {\bibfnamefont {Andrea}\
  \bibnamefont {Al{\`u}}},\ }\bibfield  {title} {\enquote {\bibinfo {title}
  {Magnetic-free non-reciprocity and isolation based on parametrically
  modulated coupled-resonator loops},}\ }\href@noop {} {\bibfield  {journal}
  {\bibinfo  {journal} {Nature Physics}\ }\textbf {\bibinfo {volume} {10}},\
  \bibinfo {pages} {923--927} (\bibinfo {year} {2014})}\BibitemShut {NoStop}%
\bibitem [{\citenamefont {Solja{\v{c}}i{\'c}}\ \emph
  {et~al.}(2003)\citenamefont {Solja{\v{c}}i{\'c}}, \citenamefont {Luo},
  \citenamefont {Joannopoulos},\ and\ \citenamefont
  {Fan}}]{soljacic2003nonlinear}%
  \BibitemOpen
  \bibfield  {author} {\bibinfo {author} {\bibfnamefont {Marin}\ \bibnamefont
  {Solja{\v{c}}i{\'c}}}, \bibinfo {author} {\bibfnamefont {Chiyan}\
  \bibnamefont {Luo}}, \bibinfo {author} {\bibfnamefont {John~D}\ \bibnamefont
  {Joannopoulos}}, \ and\ \bibinfo {author} {\bibfnamefont {Shanhui}\
  \bibnamefont {Fan}},\ }\bibfield  {title} {\enquote {\bibinfo {title}
  {Nonlinear photonic crystal microdevices for optical integration},}\
  }\href@noop {} {\bibfield  {journal} {\bibinfo  {journal} {Optics letters}\
  }\textbf {\bibinfo {volume} {28}},\ \bibinfo {pages} {637--639} (\bibinfo
  {year} {2003})}\BibitemShut {NoStop}%
\bibitem [{\citenamefont {Metelmann}\ and\ \citenamefont
  {Clerk}(2015)}]{Metelmann2015}%
  \BibitemOpen
  \bibfield  {author} {\bibinfo {author} {\bibfnamefont {A.}~\bibnamefont
  {Metelmann}}\ and\ \bibinfo {author} {\bibfnamefont {A.~A.}\ \bibnamefont
  {Clerk}},\ }\bibfield  {title} {\enquote {\bibinfo {title} {Nonreciprocal
  photon transmission and amplification via reservoir engineering},}\ }\href
  {\doibase 10.1103/PhysRevX.5.021025} {\bibfield  {journal} {\bibinfo
  {journal} {Phys. Rev. X}\ }\textbf {\bibinfo {volume} {5}},\ \bibinfo {pages}
  {021025} (\bibinfo {year} {2015})}\BibitemShut {NoStop}%
\bibitem [{\citenamefont {Ranzani}\ and\ \citenamefont
  {Aumentado}(2015)}]{ranzani2015graph}%
  \BibitemOpen
  \bibfield  {author} {\bibinfo {author} {\bibfnamefont {Leonardo}\
  \bibnamefont {Ranzani}}\ and\ \bibinfo {author} {\bibfnamefont {Jos{\'e}}\
  \bibnamefont {Aumentado}},\ }\bibfield  {title} {\enquote {\bibinfo {title}
  {Graph-based analysis of nonreciprocity in coupled-mode systems},}\
  }\href@noop {} {\bibfield  {journal} {\bibinfo  {journal} {New Journal of
  Physics}\ }\textbf {\bibinfo {volume} {17}},\ \bibinfo {pages} {023024}
  (\bibinfo {year} {2015})}\BibitemShut {NoStop}%
\bibitem [{\citenamefont {Fang}\ \emph {et~al.}(2017)\citenamefont {Fang},
  \citenamefont {Luo}, \citenamefont {Metelmann}, \citenamefont {Matheny},
  \citenamefont {Marquardt}, \citenamefont {Clerk},\ and\ \citenamefont
  {Painter}}]{fang2017generalized}%
  \BibitemOpen
  \bibfield  {author} {\bibinfo {author} {\bibfnamefont {Kejie}\ \bibnamefont
  {Fang}}, \bibinfo {author} {\bibfnamefont {Jie}\ \bibnamefont {Luo}},
  \bibinfo {author} {\bibfnamefont {Anja}\ \bibnamefont {Metelmann}}, \bibinfo
  {author} {\bibfnamefont {Matthew~H}\ \bibnamefont {Matheny}}, \bibinfo
  {author} {\bibfnamefont {Florian}\ \bibnamefont {Marquardt}}, \bibinfo
  {author} {\bibfnamefont {Aashish~A}\ \bibnamefont {Clerk}}, \ and\ \bibinfo
  {author} {\bibfnamefont {Oskar}\ \bibnamefont {Painter}},\ }\bibfield
  {title} {\enquote {\bibinfo {title} {Generalized non-reciprocity in an
  optomechanical circuit via synthetic magnetism and reservoir engineering},}\
  }\href@noop {} {\bibfield  {journal} {\bibinfo  {journal} {Nature Physics}\
  }\textbf {\bibinfo {volume} {13}},\ \bibinfo {pages} {465--471} (\bibinfo
  {year} {2017})}\BibitemShut {NoStop}%
\bibitem [{\citenamefont {Barzanjeh}\ \emph {et~al.}(2017)\citenamefont
  {Barzanjeh}, \citenamefont {Wulf}, \citenamefont {Peruzzo}, \citenamefont
  {Kalaee}, \citenamefont {Dieterle}, \citenamefont {Painter},\ and\
  \citenamefont {Fink}}]{barzanjeh2017mechanical}%
  \BibitemOpen
  \bibfield  {author} {\bibinfo {author} {\bibfnamefont {Shabir}\ \bibnamefont
  {Barzanjeh}}, \bibinfo {author} {\bibfnamefont {Matthias}\ \bibnamefont
  {Wulf}}, \bibinfo {author} {\bibfnamefont {Matilda}\ \bibnamefont {Peruzzo}},
  \bibinfo {author} {\bibfnamefont {Mahmoud}\ \bibnamefont {Kalaee}}, \bibinfo
  {author} {\bibfnamefont {PB}~\bibnamefont {Dieterle}}, \bibinfo {author}
  {\bibfnamefont {Oskar}\ \bibnamefont {Painter}}, \ and\ \bibinfo {author}
  {\bibfnamefont {Johannes~M}\ \bibnamefont {Fink}},\ }\bibfield  {title}
  {\enquote {\bibinfo {title} {Mechanical on-chip microwave circulator},}\
  }\href@noop {} {\bibfield  {journal} {\bibinfo  {journal} {Nature
  communications}\ }\textbf {\bibinfo {volume} {8}},\ \bibinfo {pages} {1--7}
  (\bibinfo {year} {2017})}\BibitemShut {NoStop}%
\bibitem [{\citenamefont {Peterson}\ \emph {et~al.}(2017)\citenamefont
  {Peterson}, \citenamefont {Lecocq}, \citenamefont {Cicak}, \citenamefont
  {Simmonds}, \citenamefont {Aumentado},\ and\ \citenamefont
  {Teufel}}]{Peterson2017}%
  \BibitemOpen
  \bibfield  {author} {\bibinfo {author} {\bibfnamefont {G.~A.}\ \bibnamefont
  {Peterson}}, \bibinfo {author} {\bibfnamefont {F.}~\bibnamefont {Lecocq}},
  \bibinfo {author} {\bibfnamefont {K.}~\bibnamefont {Cicak}}, \bibinfo
  {author} {\bibfnamefont {R.~W.}\ \bibnamefont {Simmonds}}, \bibinfo {author}
  {\bibfnamefont {J.}~\bibnamefont {Aumentado}}, \ and\ \bibinfo {author}
  {\bibfnamefont {J.~D.}\ \bibnamefont {Teufel}},\ }\bibfield  {title}
  {\enquote {\bibinfo {title} {Demonstration of efficient nonreciprocity in a
  microwave optomechanical circuit},}\ }\href {\doibase
  10.1103/PhysRevX.7.031001} {\bibfield  {journal} {\bibinfo  {journal} {Phys.
  Rev. X}\ }\textbf {\bibinfo {volume} {7}},\ \bibinfo {pages} {031001}
  (\bibinfo {year} {2017})}\BibitemShut {NoStop}%
\bibitem [{\citenamefont {Bernier}\ \emph {et~al.}(2017)\citenamefont
  {Bernier}, \citenamefont {Toth}, \citenamefont {Koottandavida}, \citenamefont
  {Ioannou}, \citenamefont {Malz}, \citenamefont {Nunnenkamp}, \citenamefont
  {Feofanov},\ and\ \citenamefont {Kippenberg}}]{bernier2017nonreciprocal}%
  \BibitemOpen
  \bibfield  {author} {\bibinfo {author} {\bibfnamefont {Nathan~Rafa{\"e}l}\
  \bibnamefont {Bernier}}, \bibinfo {author} {\bibfnamefont {Laszlo~Daniel}\
  \bibnamefont {Toth}}, \bibinfo {author} {\bibfnamefont {A}~\bibnamefont
  {Koottandavida}}, \bibinfo {author} {\bibfnamefont {Marie~Adrienne}\
  \bibnamefont {Ioannou}}, \bibinfo {author} {\bibfnamefont {Daniel}\
  \bibnamefont {Malz}}, \bibinfo {author} {\bibfnamefont {Andreas}\
  \bibnamefont {Nunnenkamp}}, \bibinfo {author} {\bibfnamefont
  {AK}~\bibnamefont {Feofanov}}, \ and\ \bibinfo {author} {\bibfnamefont
  {TJ}~\bibnamefont {Kippenberg}},\ }\bibfield  {title} {\enquote {\bibinfo
  {title} {Nonreciprocal reconfigurable microwave optomechanical circuit},}\
  }\href@noop {} {\bibfield  {journal} {\bibinfo  {journal} {Nature
  communications}\ }\textbf {\bibinfo {volume} {8}},\ \bibinfo {pages} {1--8}
  (\bibinfo {year} {2017})}\BibitemShut {NoStop}%
\bibitem [{\citenamefont {Xu}\ \emph {et~al.}(2019)\citenamefont {Xu},
  \citenamefont {Jiang}, \citenamefont {Clerk},\ and\ \citenamefont
  {Harris}}]{xu2019nonreciprocal}%
  \BibitemOpen
  \bibfield  {author} {\bibinfo {author} {\bibfnamefont {H}~\bibnamefont {Xu}},
  \bibinfo {author} {\bibfnamefont {Luyao}\ \bibnamefont {Jiang}}, \bibinfo
  {author} {\bibfnamefont {AA}~\bibnamefont {Clerk}}, \ and\ \bibinfo {author}
  {\bibfnamefont {JGE}\ \bibnamefont {Harris}},\ }\bibfield  {title} {\enquote
  {\bibinfo {title} {Nonreciprocal control and cooling of phonon modes in an
  optomechanical system},}\ }\href@noop {} {\bibfield  {journal} {\bibinfo
  {journal} {Nature}\ }\textbf {\bibinfo {volume} {568}},\ \bibinfo {pages}
  {65--69} (\bibinfo {year} {2019})}\BibitemShut {NoStop}%
\bibitem [{\citenamefont {Porras}\ and\ \citenamefont
  {Fern\'andez-Lorenzo}(2019)}]{Porras2019}%
  \BibitemOpen
  \bibfield  {author} {\bibinfo {author} {\bibfnamefont {Diego}\ \bibnamefont
  {Porras}}\ and\ \bibinfo {author} {\bibfnamefont {Samuel}\ \bibnamefont
  {Fern\'andez-Lorenzo}},\ }\bibfield  {title} {\enquote {\bibinfo {title}
  {Topological amplification in photonic lattices},}\ }\href {\doibase
  10.1103/PhysRevLett.122.143901} {\bibfield  {journal} {\bibinfo  {journal}
  {Phys. Rev. Lett.}\ }\textbf {\bibinfo {volume} {122}},\ \bibinfo {pages}
  {143901} (\bibinfo {year} {2019})}\BibitemShut {NoStop}%
\bibitem [{\citenamefont {Ruesink}\ \emph {et~al.}(2016)\citenamefont
  {Ruesink}, \citenamefont {Miri}, \citenamefont {Alu},\ and\ \citenamefont
  {Verhagen}}]{ruesink2016nonreciprocity}%
  \BibitemOpen
  \bibfield  {author} {\bibinfo {author} {\bibfnamefont {Freek}\ \bibnamefont
  {Ruesink}}, \bibinfo {author} {\bibfnamefont {Mohammad-Ali}\ \bibnamefont
  {Miri}}, \bibinfo {author} {\bibfnamefont {Andrea}\ \bibnamefont {Alu}}, \
  and\ \bibinfo {author} {\bibfnamefont {Ewold}\ \bibnamefont {Verhagen}},\
  }\bibfield  {title} {\enquote {\bibinfo {title} {Nonreciprocity and
  magnetic-free isolation based on optomechanical interactions},}\ }\href@noop
  {} {\bibfield  {journal} {\bibinfo  {journal} {Nature communications}\
  }\textbf {\bibinfo {volume} {7}},\ \bibinfo {pages} {1--8} (\bibinfo {year}
  {2016})}\BibitemShut {NoStop}%
\bibitem [{\citenamefont {Barzanjeh}\ \emph {et~al.}(2018)\citenamefont
  {Barzanjeh}, \citenamefont {Aquilina},\ and\ \citenamefont
  {Xuereb}}]{Barzanjeh2018}%
  \BibitemOpen
  \bibfield  {author} {\bibinfo {author} {\bibfnamefont {Shabir}\ \bibnamefont
  {Barzanjeh}}, \bibinfo {author} {\bibfnamefont {Matteo}\ \bibnamefont
  {Aquilina}}, \ and\ \bibinfo {author} {\bibfnamefont {Andr\'e}\ \bibnamefont
  {Xuereb}},\ }\bibfield  {title} {\enquote {\bibinfo {title} {Manipulating the
  flow of thermal noise in quantum devices},}\ }\href {\doibase
  10.1103/PhysRevLett.120.060601} {\bibfield  {journal} {\bibinfo  {journal}
  {Phys. Rev. Lett.}\ }\textbf {\bibinfo {volume} {120}},\ \bibinfo {pages}
  {060601} (\bibinfo {year} {2018})}\BibitemShut {NoStop}%
\bibitem [{\citenamefont {McDonald}\ \emph {et~al.}(2018)\citenamefont
  {McDonald}, \citenamefont {Pereg-Barnea},\ and\ \citenamefont
  {Clerk}}]{McDonald2018phase}%
  \BibitemOpen
  \bibfield  {author} {\bibinfo {author} {\bibfnamefont {A.}~\bibnamefont
  {McDonald}}, \bibinfo {author} {\bibfnamefont {T.}~\bibnamefont
  {Pereg-Barnea}}, \ and\ \bibinfo {author} {\bibfnamefont {A.~A.}\
  \bibnamefont {Clerk}},\ }\bibfield  {title} {\enquote {\bibinfo {title}
  {Phase-dependent chiral transport and effective non-hermitian dynamics in a
  bosonic kitaev-majorana chain},}\ }\href {\doibase 10.1103/PhysRevX.8.041031}
  {\bibfield  {journal} {\bibinfo  {journal} {Phys. Rev. X}\ }\textbf {\bibinfo
  {volume} {8}},\ \bibinfo {pages} {041031} (\bibinfo {year}
  {2018})}\BibitemShut {NoStop}%
\bibitem [{\citenamefont {Wanjura}\ \emph {et~al.}(2020)\citenamefont
  {Wanjura}, \citenamefont {Brunelli},\ and\ \citenamefont
  {Nunnenkamp}}]{Wanjura2019}%
  \BibitemOpen
  \bibfield  {author} {\bibinfo {author} {\bibfnamefont {Clara~C}\ \bibnamefont
  {Wanjura}}, \bibinfo {author} {\bibfnamefont {Matteo}\ \bibnamefont
  {Brunelli}}, \ and\ \bibinfo {author} {\bibfnamefont {Andreas}\ \bibnamefont
  {Nunnenkamp}},\ }\bibfield  {title} {\enquote {\bibinfo {title} {Topological
  framework for directional amplification in driven-dissipative cavity
  arrays},}\ }\href@noop {} {\bibfield  {journal} {\bibinfo  {journal} {Nature
  communications}\ }\textbf {\bibinfo {volume} {11}},\ \bibinfo {pages} {3149}
  (\bibinfo {year} {2020})}\BibitemShut {NoStop}%
\bibitem [{\citenamefont {Wang}\ \emph {et~al.}(2019)\citenamefont {Wang},
  \citenamefont {Rao}, \citenamefont {Yang}, \citenamefont {Xu}, \citenamefont
  {Gui}, \citenamefont {Yao}, \citenamefont {You},\ and\ \citenamefont
  {Hu}}]{Wang2019Nonreciprocity}%
  \BibitemOpen
  \bibfield  {author} {\bibinfo {author} {\bibfnamefont {Yi-Pu}\ \bibnamefont
  {Wang}}, \bibinfo {author} {\bibfnamefont {J.~W.}\ \bibnamefont {Rao}},
  \bibinfo {author} {\bibfnamefont {Y.}~\bibnamefont {Yang}}, \bibinfo {author}
  {\bibfnamefont {Peng-Chao}\ \bibnamefont {Xu}}, \bibinfo {author}
  {\bibfnamefont {Y.~S.}\ \bibnamefont {Gui}}, \bibinfo {author} {\bibfnamefont
  {B.~M.}\ \bibnamefont {Yao}}, \bibinfo {author} {\bibfnamefont {J.~Q.}\
  \bibnamefont {You}}, \ and\ \bibinfo {author} {\bibfnamefont {C.-M.}\
  \bibnamefont {Hu}},\ }\bibfield  {title} {\enquote {\bibinfo {title}
  {Nonreciprocity and unidirectional invisibility in cavity magnonics},}\
  }\href {\doibase 10.1103/PhysRevLett.123.127202} {\bibfield  {journal}
  {\bibinfo  {journal} {Phys. Rev. Lett.}\ }\textbf {\bibinfo {volume} {123}},\
  \bibinfo {pages} {127202} (\bibinfo {year} {2019})}\BibitemShut {NoStop}%
\bibitem [{\citenamefont {Malz}\ \emph {et~al.}(2018)\citenamefont {Malz},
  \citenamefont {T\'oth}, \citenamefont {Bernier}, \citenamefont {Feofanov},
  \citenamefont {Kippenberg},\ and\ \citenamefont {Nunnenkamp}}]{Malz2018}%
  \BibitemOpen
  \bibfield  {author} {\bibinfo {author} {\bibfnamefont {Daniel}\ \bibnamefont
  {Malz}}, \bibinfo {author} {\bibfnamefont {L\'aszl\'o~D.}\ \bibnamefont
  {T\'oth}}, \bibinfo {author} {\bibfnamefont {Nathan~R.}\ \bibnamefont
  {Bernier}}, \bibinfo {author} {\bibfnamefont {Alexey~K.}\ \bibnamefont
  {Feofanov}}, \bibinfo {author} {\bibfnamefont {Tobias~J.}\ \bibnamefont
  {Kippenberg}}, \ and\ \bibinfo {author} {\bibfnamefont {Andreas}\
  \bibnamefont {Nunnenkamp}},\ }\bibfield  {title} {\enquote {\bibinfo {title}
  {Quantum-limited directional amplifiers with optomechanics},}\ }\href
  {\doibase 10.1103/PhysRevLett.120.023601} {\bibfield  {journal} {\bibinfo
  {journal} {Phys. Rev. Lett.}\ }\textbf {\bibinfo {volume} {120}},\ \bibinfo
  {pages} {023601} (\bibinfo {year} {2018})}\BibitemShut {NoStop}%
\bibitem [{\citenamefont {Mercier~de L\'epinay}\ \emph
  {et~al.}(2019)\citenamefont {Mercier~de L\'epinay}, \citenamefont
  {Damsk\"agg}, \citenamefont {Ockeloen-Korppi},\ and\ \citenamefont
  {Sillanp\"a\"a}}]{Lepinay2019}%
  \BibitemOpen
  \bibfield  {author} {\bibinfo {author} {\bibfnamefont {Laure}\ \bibnamefont
  {Mercier~de L\'epinay}}, \bibinfo {author} {\bibfnamefont {Erno}\
  \bibnamefont {Damsk\"agg}}, \bibinfo {author} {\bibfnamefont {Caspar~F.}\
  \bibnamefont {Ockeloen-Korppi}}, \ and\ \bibinfo {author} {\bibfnamefont
  {Mika~A.}\ \bibnamefont {Sillanp\"a\"a}},\ }\bibfield  {title} {\enquote
  {\bibinfo {title} {Realization of directional amplification in a microwave
  optomechanical device},}\ }\href {\doibase 10.1103/PhysRevApplied.11.034027}
  {\bibfield  {journal} {\bibinfo  {journal} {Phys. Rev. Applied}\ }\textbf
  {\bibinfo {volume} {11}},\ \bibinfo {pages} {034027} (\bibinfo {year}
  {2019})}\BibitemShut {NoStop}%
\bibitem [{\citenamefont {Zhang}\ \emph {et~al.}(2020)\citenamefont {Zhang},
  \citenamefont {Yang},\ and\ \citenamefont {Fang}}]{Zhang2020correspondence}%
  \BibitemOpen
  \bibfield  {author} {\bibinfo {author} {\bibfnamefont {Kai}\ \bibnamefont
  {Zhang}}, \bibinfo {author} {\bibfnamefont {Zhesen}\ \bibnamefont {Yang}}, \
  and\ \bibinfo {author} {\bibfnamefont {Chen}\ \bibnamefont {Fang}},\
  }\bibfield  {title} {\enquote {\bibinfo {title} {Correspondence between
  winding numbers and skin modes in non-hermitian systems},}\ }\href {\doibase
  10.1103/PhysRevLett.125.126402} {\bibfield  {journal} {\bibinfo  {journal}
  {Phys. Rev. Lett.}\ }\textbf {\bibinfo {volume} {125}},\ \bibinfo {pages}
  {126402} (\bibinfo {year} {2020})}\BibitemShut {NoStop}%
\bibitem [{\citenamefont {Okuma}\ \emph {et~al.}(2020)\citenamefont {Okuma},
  \citenamefont {Kawabata}, \citenamefont {Shiozaki},\ and\ \citenamefont
  {Sato}}]{Okuma2020}%
  \BibitemOpen
  \bibfield  {author} {\bibinfo {author} {\bibfnamefont {Nobuyuki}\
  \bibnamefont {Okuma}}, \bibinfo {author} {\bibfnamefont {Kohei}\ \bibnamefont
  {Kawabata}}, \bibinfo {author} {\bibfnamefont {Ken}\ \bibnamefont
  {Shiozaki}}, \ and\ \bibinfo {author} {\bibfnamefont {Masatoshi}\
  \bibnamefont {Sato}},\ }\bibfield  {title} {\enquote {\bibinfo {title}
  {Topological origin of non-hermitian skin effects},}\ }\href {\doibase
  10.1103/PhysRevLett.124.086801} {\bibfield  {journal} {\bibinfo  {journal}
  {Phys. Rev. Lett.}\ }\textbf {\bibinfo {volume} {124}},\ \bibinfo {pages}
  {086801} (\bibinfo {year} {2020})}\BibitemShut {NoStop}%
\bibitem [{sup()}]{supplemental}%
  \BibitemOpen
  \href@noop {} {}\bibinfo {howpublished} {See the Supplemental
  Material.}\BibitemShut {Stop}%
\bibitem [{\citenamefont {Aspelmeyer}\ \emph {et~al.}(2014)\citenamefont
  {Aspelmeyer}, \citenamefont {Kippenberg},\ and\ \citenamefont
  {Marquardt}}]{Aspelmeyer2014RMP}%
  \BibitemOpen
  \bibfield  {author} {\bibinfo {author} {\bibfnamefont {Markus}\ \bibnamefont
  {Aspelmeyer}}, \bibinfo {author} {\bibfnamefont {Tobias~J.}\ \bibnamefont
  {Kippenberg}}, \ and\ \bibinfo {author} {\bibfnamefont {Florian}\
  \bibnamefont {Marquardt}},\ }\bibfield  {title} {\enquote {\bibinfo {title}
  {Cavity optomechanics},}\ }\href {\doibase 10.1103/RevModPhys.86.1391}
  {\bibfield  {journal} {\bibinfo  {journal} {Rev. Mod. Phys.}\ }\textbf
  {\bibinfo {volume} {86}},\ \bibinfo {pages} {1391--1452} (\bibinfo {year}
  {2014})}\BibitemShut {NoStop}%
\bibitem [{\citenamefont {Fitzpatrick}\ \emph {et~al.}(2017)\citenamefont
  {Fitzpatrick}, \citenamefont {Sundaresan}, \citenamefont {Li}, \citenamefont
  {Koch},\ and\ \citenamefont {Houck}}]{fitzpatrick2017observation}%
  \BibitemOpen
  \bibfield  {author} {\bibinfo {author} {\bibfnamefont {Mattias}\ \bibnamefont
  {Fitzpatrick}}, \bibinfo {author} {\bibfnamefont {Neereja~M}\ \bibnamefont
  {Sundaresan}}, \bibinfo {author} {\bibfnamefont {Andy~CY}\ \bibnamefont
  {Li}}, \bibinfo {author} {\bibfnamefont {Jens}\ \bibnamefont {Koch}}, \ and\
  \bibinfo {author} {\bibfnamefont {Andrew~A}\ \bibnamefont {Houck}},\
  }\bibfield  {title} {\enquote {\bibinfo {title} {Observation of a dissipative
  phase transition in a one-dimensional circuit qed lattice},}\ }\href@noop {}
  {\bibfield  {journal} {\bibinfo  {journal} {Physical Review X}\ }\textbf
  {\bibinfo {volume} {7}},\ \bibinfo {pages} {011016} (\bibinfo {year}
  {2017})}\BibitemShut {NoStop}%
\bibitem [{Note2()}]{Note2}%
  \BibitemOpen
  \bibinfo {note} {It can be viewed as a boson counterpart of the fermion
  damping matrix\cite {Song2019}.}\BibitemShut {Stop}%
\bibitem [{\citenamefont {Clerk}\ \emph {et~al.}(2010)\citenamefont {Clerk},
  \citenamefont {Devoret}, \citenamefont {Girvin}, \citenamefont {Marquardt},\
  and\ \citenamefont {Schoelkopf}}]{Clerk2010}%
  \BibitemOpen
  \bibfield  {author} {\bibinfo {author} {\bibfnamefont {A.~A.}\ \bibnamefont
  {Clerk}}, \bibinfo {author} {\bibfnamefont {M.~H.}\ \bibnamefont {Devoret}},
  \bibinfo {author} {\bibfnamefont {S.~M.}\ \bibnamefont {Girvin}}, \bibinfo
  {author} {\bibfnamefont {Florian}\ \bibnamefont {Marquardt}}, \ and\ \bibinfo
  {author} {\bibfnamefont {R.~J.}\ \bibnamefont {Schoelkopf}},\ }\bibfield
  {title} {\enquote {\bibinfo {title} {Introduction to quantum noise,
  measurement, and amplification},}\ }\href {\doibase
  10.1103/RevModPhys.82.1155} {\bibfield  {journal} {\bibinfo  {journal} {Rev.
  Mod. Phys.}\ }\textbf {\bibinfo {volume} {82}},\ \bibinfo {pages}
  {1155--1208} (\bibinfo {year} {2010})}\BibitemShut {NoStop}%
\bibitem [{\citenamefont {Song}\ \emph
  {et~al.}(2019{\natexlab{b}})\citenamefont {Song}, \citenamefont {Yao},\ and\
  \citenamefont {Wang}}]{Song2019}%
  \BibitemOpen
  \bibfield  {author} {\bibinfo {author} {\bibfnamefont {Fei}\ \bibnamefont
  {Song}}, \bibinfo {author} {\bibfnamefont {Shunyu}\ \bibnamefont {Yao}}, \
  and\ \bibinfo {author} {\bibfnamefont {Zhong}\ \bibnamefont {Wang}},\
  }\bibfield  {title} {\enquote {\bibinfo {title} {Non-hermitian skin effect
  and chiral damping in open quantum systems},}\ }\href {\doibase
  10.1103/PhysRevLett.123.170401} {\bibfield  {journal} {\bibinfo  {journal}
  {Phys. Rev. Lett.}\ }\textbf {\bibinfo {volume} {123}},\ \bibinfo {pages}
  {170401} (\bibinfo {year} {2019}{\natexlab{b}})}\BibitemShut {NoStop}%
\end{thebibliography}%

\clearpage

{\bf Supplemental Material}  

\vspace{7mm}

\section{Derivation of the GBZ-based Green's function formula}

We now derive Eq. (8) of the main article: \bea
G_{ij}(\omega)=\int_{\text{GBZ}}\frac{d\beta}{2\pi i\beta} \frac{\beta^{i-j}}{\omega-h(\beta)},  \label{formula}
\eea where GBZ stands for the generalized Brillouin zone\cite{yao2018edge,Yokomizo2019}.

\begin{figure}
\includegraphics[width=4.2cm, height=4cm]{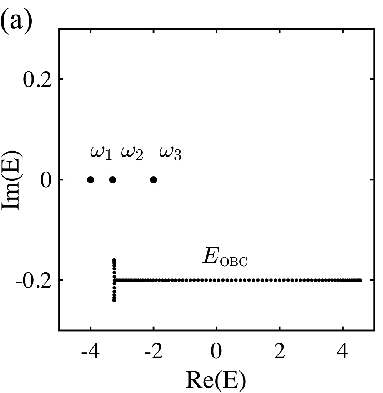}
\includegraphics[width=4.2cm, height=4cm]{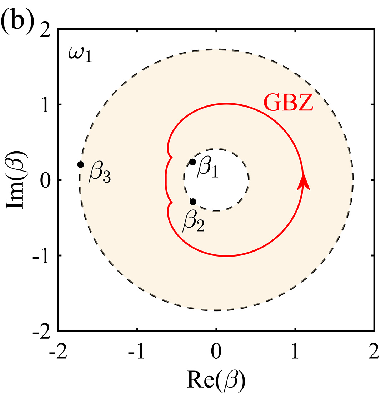}
\includegraphics[width=4.2cm, height=4cm]{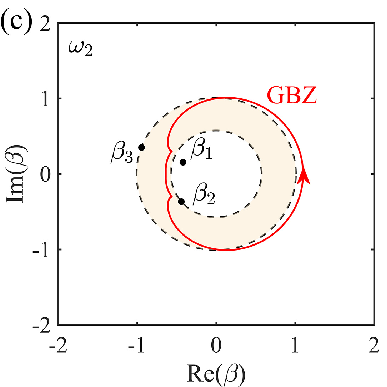}
\includegraphics[width=4.2cm, height=4cm]{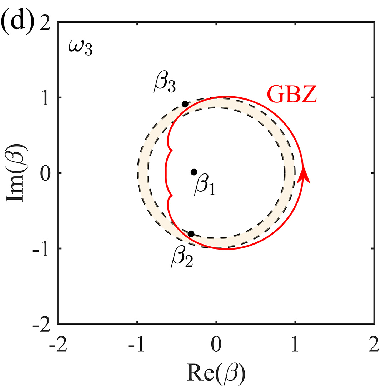}
\caption{GBZ and roots of $h(\beta)=\omega$. ({\bf a}) The OBC energy spectra $E_\text{OBC}$, and frequencies $\omega_{1,2,3}$ used in (b,c,d). Values of $t_1,t_2,\gamma$ are the same as in Fig. 2 of the main article, and $\kappa=-0.2$ (Stability requires that the imaginary parts of energies are negative, which is satisfied when $\kappa<-0.04$). ({\bf b,c,d}) The roots $\beta_{1,2,3}$ for $\omega_1=-4, \omega_2=-3.3$, and $\omega_3=-2$. The fourth root $\beta_4$ is not shown for being outside this region.  }
\label{fig3}
\end{figure}

For a given Laurent polynomial $f(\beta)=\sum_j f_j\beta^j$, one can define a matrix $T(f)$ (known as a Toeplitz matrix) whose elements are $T_{jk}(f)=f_{k-j}$, or $T_{jk}(f)=\int_{|\beta|=R}\frac{d\beta}{2\pi i\beta}\beta^{j-k}f(\beta)$ for an arbitrary radius $R$. The rank of $T$ is the chain length in our work, i.e. $j,k=1,\cdots,L$. We have $\omega-H=T(\omega-h(\beta))$ by definition, and our task is to calculate its inverse, $G$. To this end, we use a product identity $T(f)T(g)=T(fg)$ for two Laurent polynomials $f$ and $g$. This can be proved by $\sum_k T_{ik}(f)T_{kj}(g) =\sum_k f_{k-i}g_{j-k} =(fg)_{j-i}=T_{ij}(fg)$ (There are some corrections near the boundaries $i,j=1$ or $L$, which will not be our focus). As a corollary, we have $T(f)T(f^{-1})=T(1)$ and $[T(f)]^{-1}=T(f^{-1})$. We may formally take $f=\omega-h(\beta)$, then $G=T(\frac{1}{\omega-h(\beta)})$, and consequently \bea
G_{ij}=\int_{|\beta|=R}\frac{d\beta}{2\pi i\beta} \frac{\beta^{i-j}}{\omega-h(\beta)}.
\label{R}
\eea
This means expanding $\frac{1}{\omega-h(\beta)}$ as a Laurent series and the coefficients are $G_{ij}$. However, as a mathematical fact, the Laurent series depends on $R$. In fact, in $T(f)T(f^{-1})=T(1)$, the proper Laurent series for $f^{-1}$ should be obtained as follows. We find a smooth interpolation $f_t(\beta)$ between $f_{t=1}(\beta)=f(\beta)$ and the trivial polynomial $f_{t=0}(\beta)=1$, for which we know that $T(f_{t=0})T(f_{t=0}^{-1})=T(1)$ is trivially true. The proper Laurent series for $f_{t=1}^{-1}(\beta)=f^{-1}(\beta)$ is then obtained from the interpolation $f_t^{-1}(\beta)$. The smoothness of interpolation means that $f_t(\beta)\neq 0$ on the circle $|\beta|=R$. The existence of such an interpolation requires that the phase winding number $\frac{1}{2\pi}\int_{|\beta|=R} d\arg[f(\beta)]=0$ because, as a topological invariant, the winding number $\frac{1}{2\pi}\int_{|\beta|=R} d\arg[f_t(\beta)]$ stays constant as $t$ varies, and $f_{t=0}$ apparently has a vanishing winding number. To implement this topological condition, let us factorize
\bea
\omega-h(\beta) = \frac{ c_M\prod_{n=1}^{2M}(\beta-\beta_n)}{\beta^M}, \label{factors}
\eea
where $\beta_{n=1,\cdots,2M}$ stand for the roots of $\omega-h(\beta)=0$ and $c_M$ is the coefficient of $\beta^M$. It follows that the vanishing of winding number requires taking $R$ within $|\beta_M|<R<|\beta_{M+1}|$ [the colored region in Fig. \ref{fig3}(b,c,d)]. Thus the circle $|\beta|=R$ encloses $\beta_1,\cdots,\beta_M$. As discussed before, the GBZ encloses the same roots, and therefore we can replace the integration contour in Eq. (\ref{R}) by the GBZ, resulting in Eq. (\ref{formula}). Notably, although the eligible $R$ region in Eq. (\ref{R}) varies with $\omega$ [colored in Fig. \ref{fig3}(b,c,d)], Eq. (\ref{formula}) remains valid with the same GBZ for all $\omega$.

\section{Explicit expressions of $G_{ij}$ in terms of roots}

In the main text, we have presented the GBZ-based Green's function formula, reproduced as Eq. (\ref{formula}) in this Supplemental Material. It can be readily simplified by the residue theorem. The calculation is rather straightforward; nevertheless, we provide the explicit results for the sake of completeness. The denominator of the integrand can be written as $\omega-h(\beta) = \frac{ c_M\prod_{n=1}^{2M}(\beta-\beta_n)}{\beta^M}$, where $c_M$ is the coefficient of $\beta^M$ and $\beta_{n=1,2,\cdots,2M}$ are the roots of $h(\beta)=\omega$. The integral becomes
\bea
G_{ij}(\omega)=\int_{\text{GBZ}}\frac{d\beta}{2\pi i } \frac{\beta^{i-j+M-1}}{c_M\prod_{n=1}^{2M} (\beta-\beta_n)}.
\eea
As shown in the main text, when the roots are ordered as $|\beta_1|\leq|\beta_2|\leq\cdots\leq|\beta_{2M}|$,  $\beta_1,\cdots,\beta_M$ are enclosed by the GBZ. Therefore, for $i-j+M-1\geq 0$, the residue theorem tells us that
\bea
G_{ij}(\omega)=\sum_{n=1}^{M}\frac{\beta_n^{i-j+M-1}}{c_M\prod_{k\neq n} (\beta_n-\beta_k)}.
\eea
When $i-j+M-1<0$, the result is
\bea
G_{ij}(\omega)=-\sum_{n=M+1}^{2M}\frac{\beta_n^{i-j+M-1}}{c_M\prod_{k\neq n} (\beta_n-\beta_k)}.
\eea
Taking $i-j=L-1$ and $-(L-1)$, we immediately see that $G_{L1}\sim(\beta_M)^L$ and $G_{1L}\sim (\beta_{M+1})^{-L}$ for large $L$.

\begin{figure*}
\includegraphics[width=4.2cm, height=4cm]{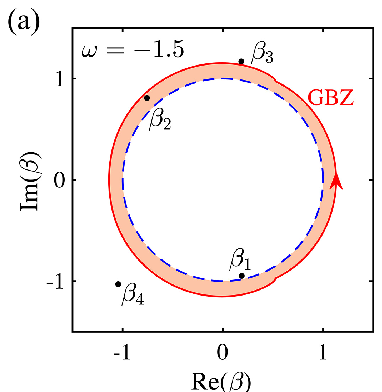}
\includegraphics[width=5cm, height=4cm]{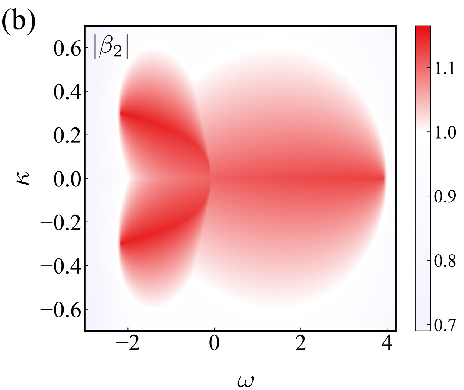}
\includegraphics[width=4.2cm, height=4cm]{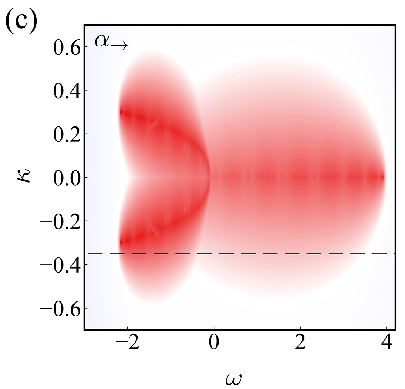}
\includegraphics[width=4.2cm, height=4cm]{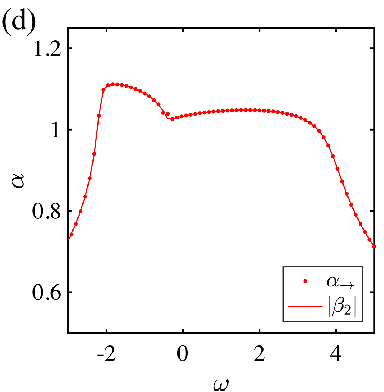}
\caption{Directional amplification with a frequency-independent directionality. Parameter values are $t_1=1, t_2=1, \gamma=0.6$. ({\bf a}) GBZ (red solid loop) and BZ (blue dashed circle). The roots $\beta_{1,2,3,4}$ of $\omega-h(\beta)=0$ are shown for $\kappa=-0.35$ and $\omega=-1.5$. ({\bf b}) $|\beta_2|$ as a function of $\kappa$ and $\omega$. ({\bf c}) $\alpha_\rw$ as a function of $\kappa$ and $\omega$, which is in agreement with the theory of (b).  ({\bf d}) Detailed comparison of $\alpha_\rw$ and $|\beta_2|$ along the cut $\kappa=-0.35$ [dashed line in (c)].}
\label{oneway}
\end{figure*}

\section{\label{sec:level2}Amplification with frequency-independent directionality}

In the main article, we focused on the case that the amplification direction depends on the frequency. As mentioned there, the same model also supports directional amplification with frequency-independent directionality. In fact, this can be achieved by taking different parameter values in the same model; an example is given in Fig. \ref{oneway}. For this choice of parameter values, the GBZ has no intersection with BZ, and therefore the amplification has to be unidirectional within the entire bandwidth.

\begin{figure}
\includegraphics[width=4.2cm, height=4cm]{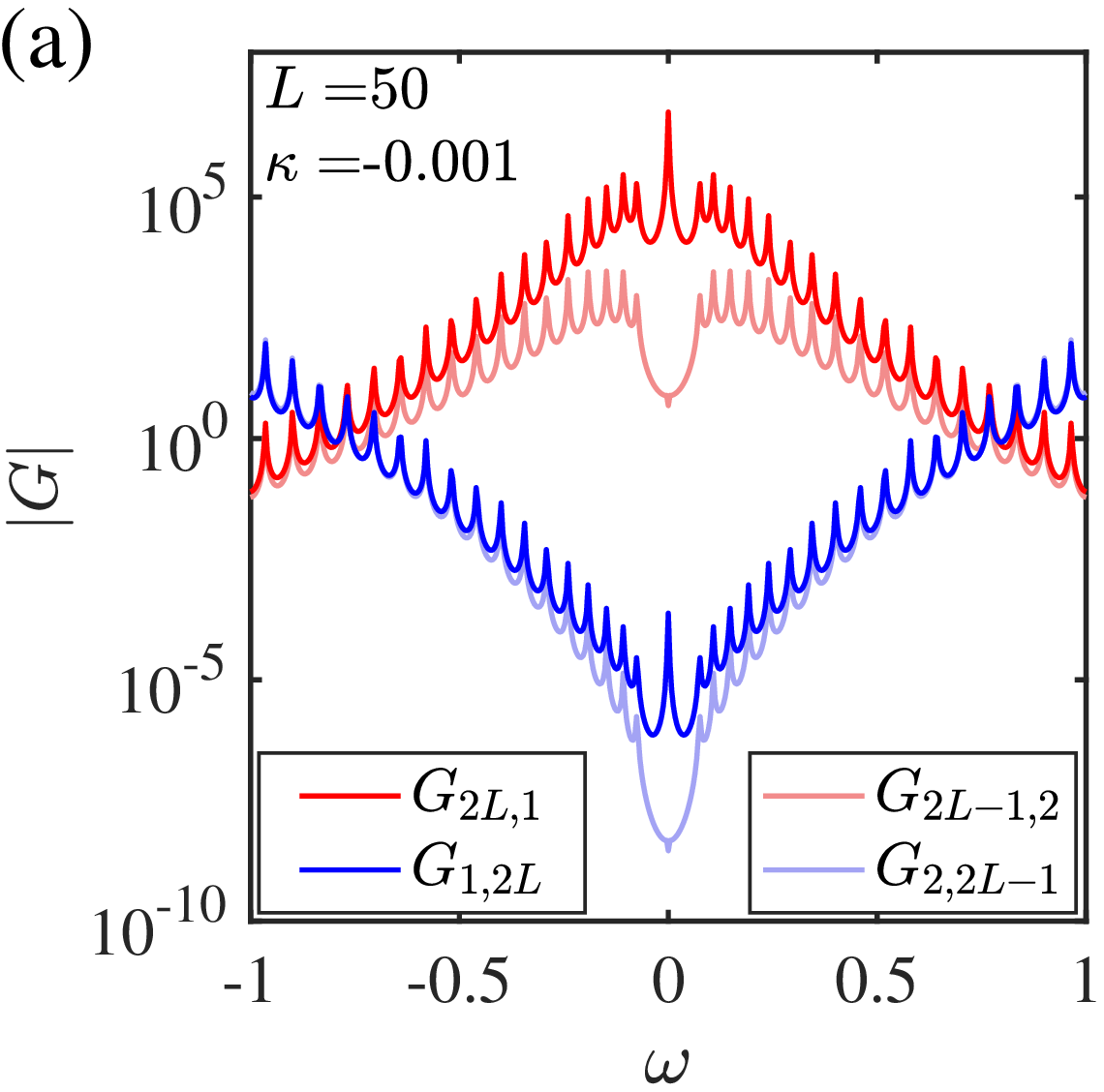}
\includegraphics[width=4.2cm, height=4cm]{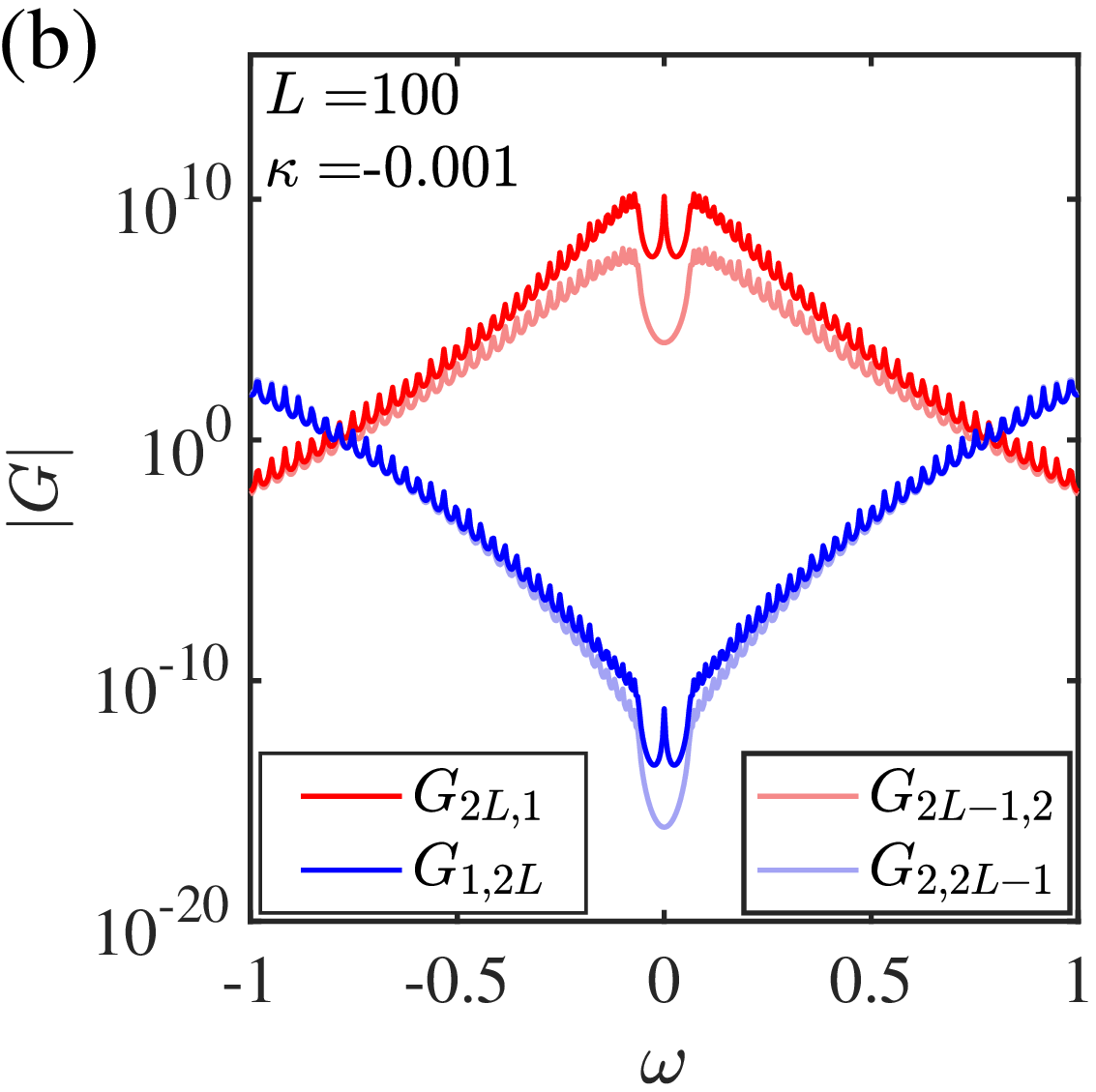}
\includegraphics[width=4.2cm, height=4cm]{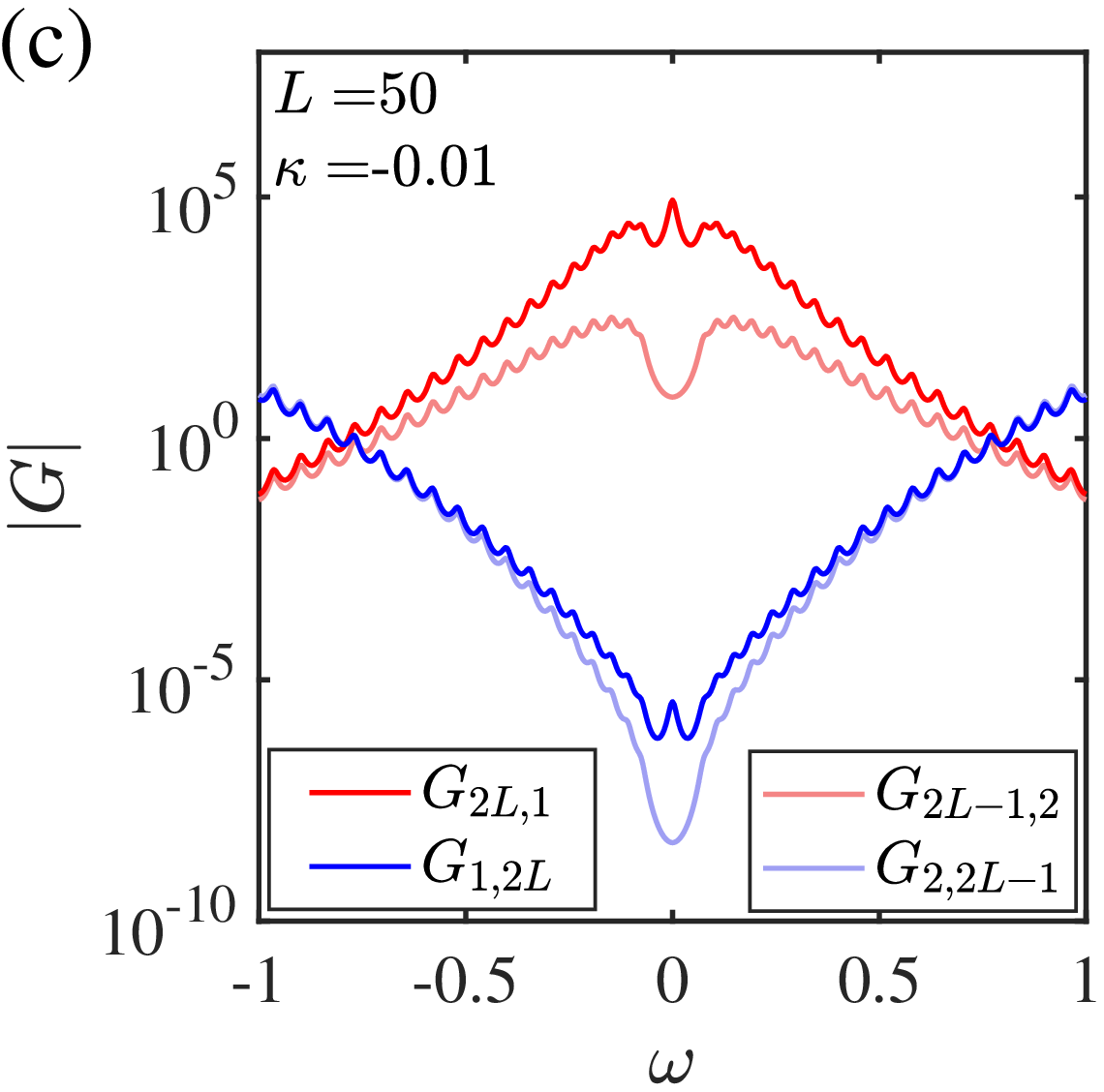}
\includegraphics[width=4.2cm, height=4cm]{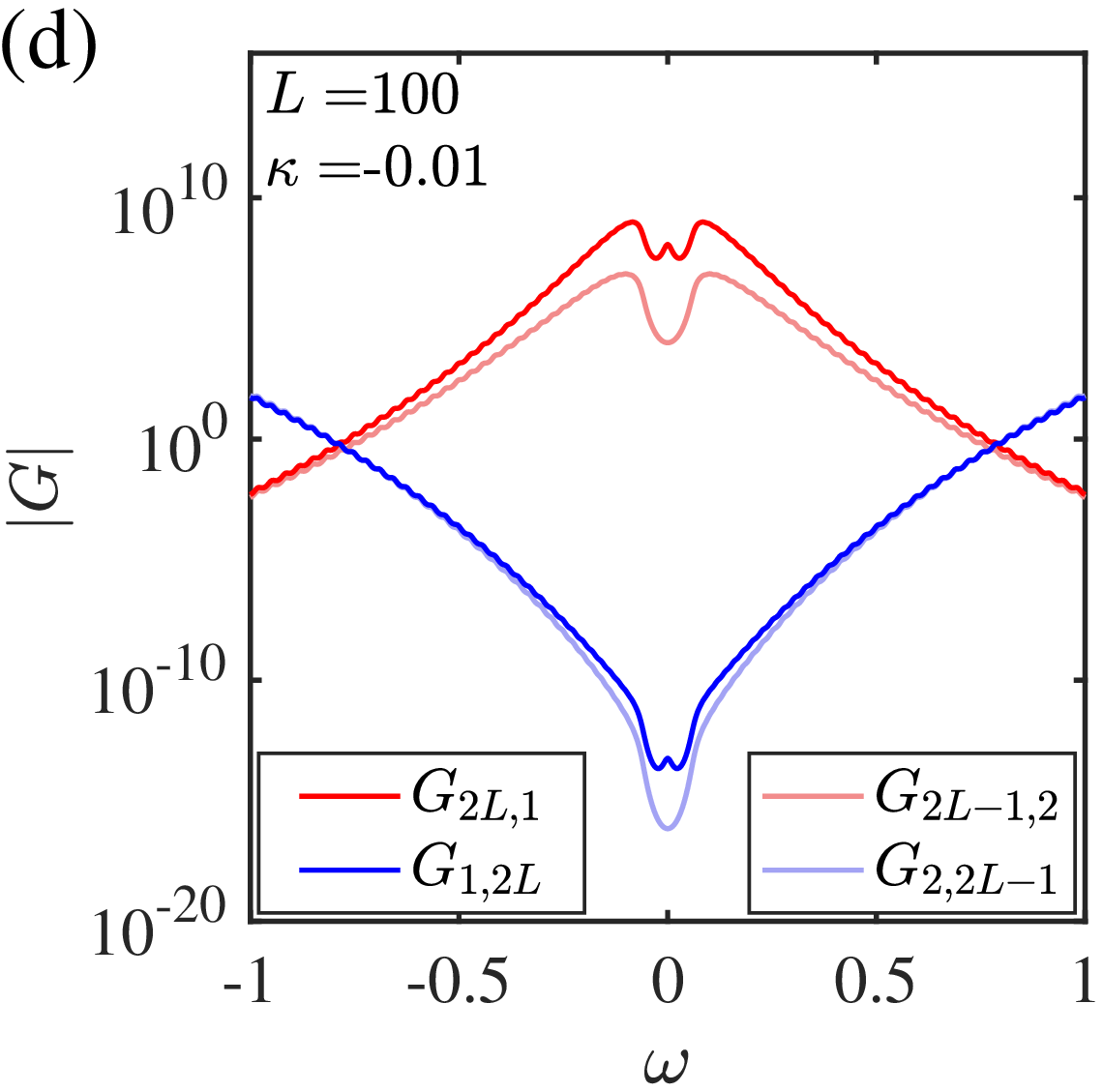}
\caption{ Green's function $G_{ij}$ ($i,j=1,2,2L-1,2L$) of the two-band model Eq. (\ref{twobandH}). Parameter values are $t_1=1.2$, $t_2=1$, $t_3=0.2$, $\gamma=0.3$. ({\bf a}) $\kappa=-0.001$, $L=50$.  ({\bf b}) $\kappa=-0.001$, $L=100$. ({\bf c}) $\kappa=-0.01$, $L=50$. ({\bf d}) $\kappa=-0.01$, $L=100$.  }
\label{zeromode}
\end{figure}

\begin{figure}
\includegraphics[width=4.2cm, height=4cm]{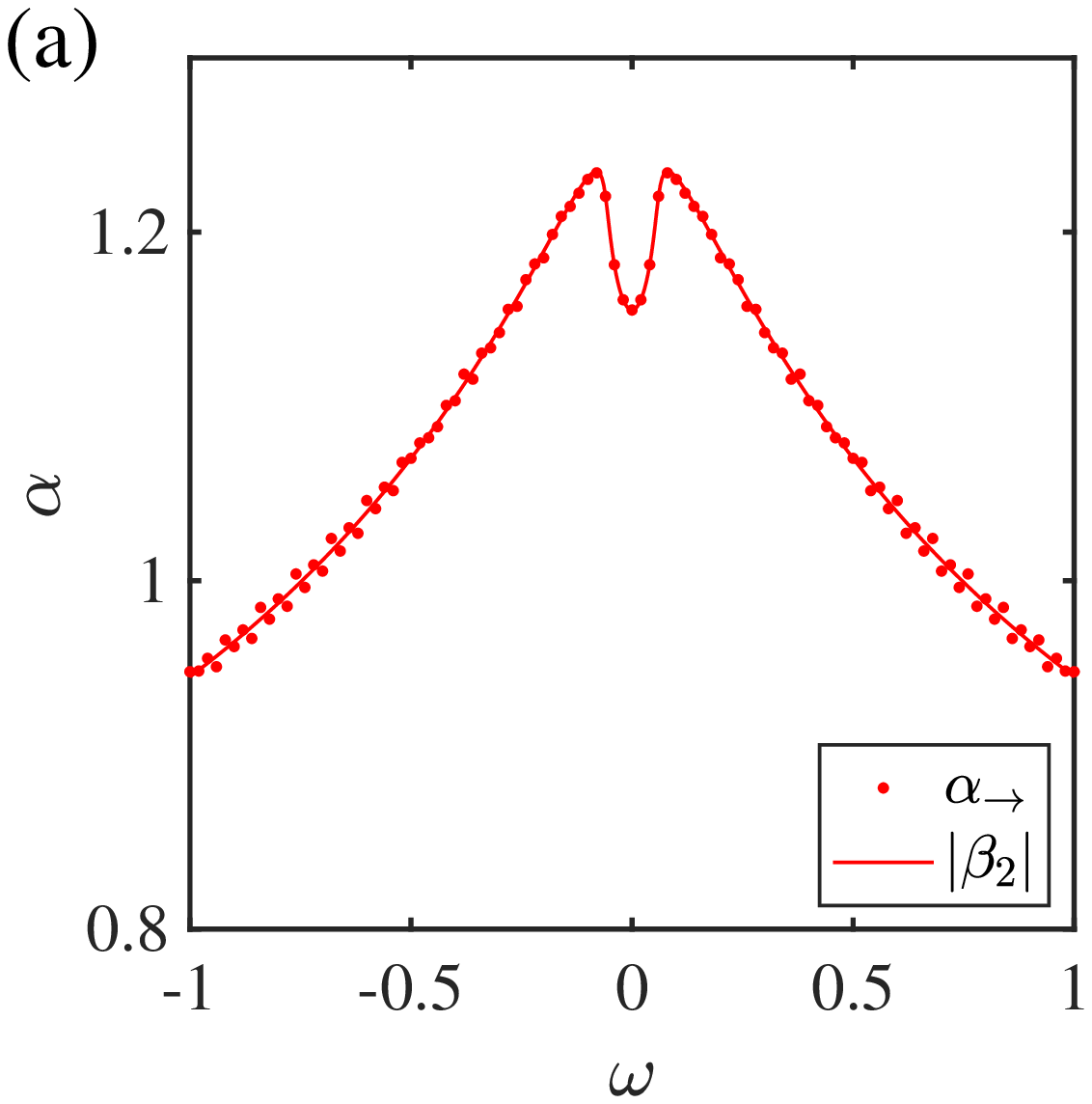}
\includegraphics[width=4.2cm, height=4cm]{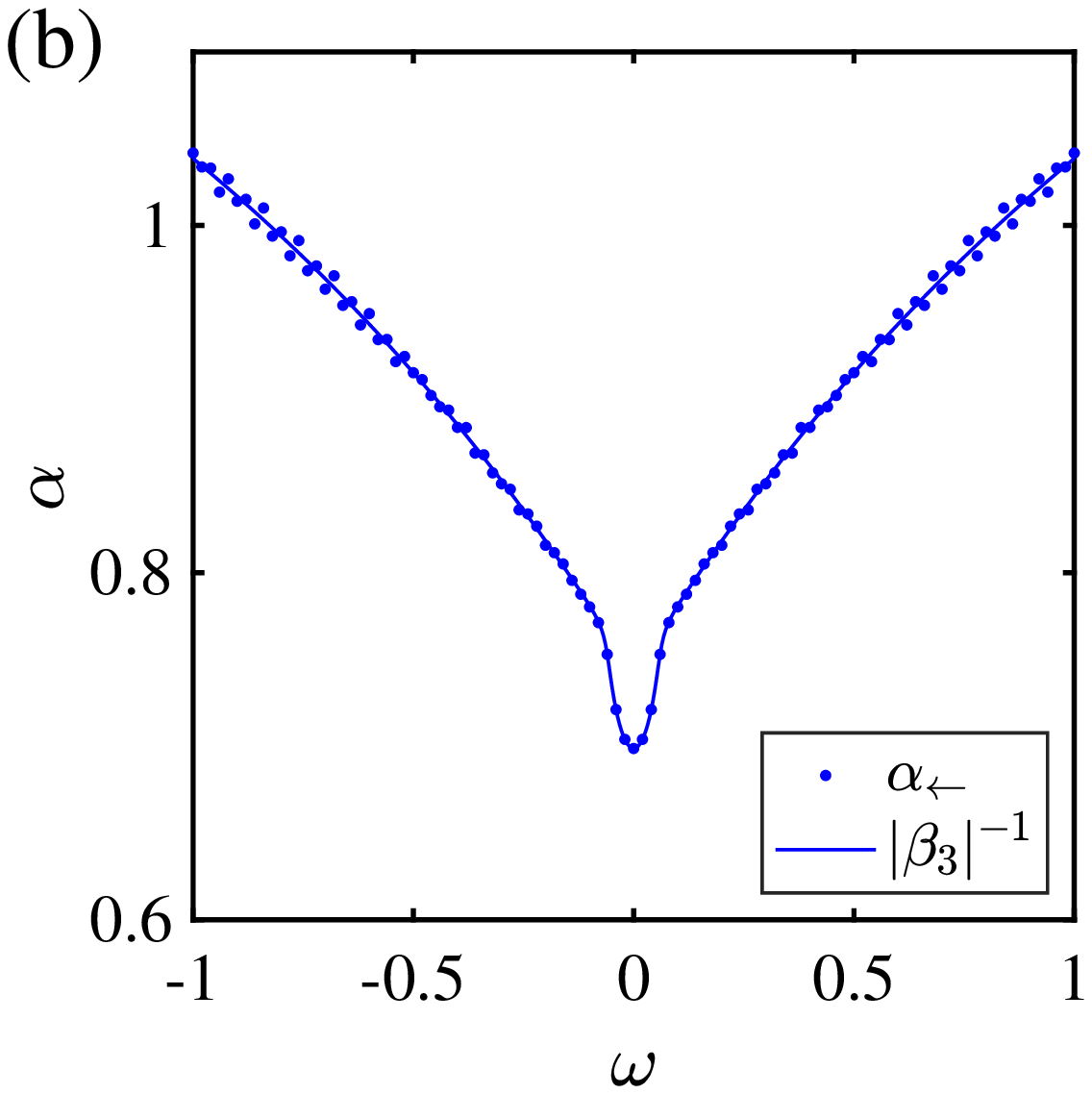}
\caption{ Comparison of our formula and brute-force numerical results for the two-band model. ({\bf a}) $\alpha_\rw$ and $|\beta_2|$. ({\bf b}) $\alpha_\lw$ and $|\beta_3|^{-1}$.  Parameter values are $t_1=1.2$, $t_2=1$, $t_3=0.2$, $\gamma=0.3$, $\kappa=-0.01$.  }
\label{twoband}
\end{figure}

\section{\label{sec:level1}Two-band model and effect of zero modes}

In the main article, we considered a single-band model without topological edge mode, for which the directional amplification comes solely from the continuous band. Here, for completeness, we consider a model with topological edge modes. We show that the zero modes have a visible contribution within a small frequency window when the chain is short, and its effect diminishes as the chain becomes longer. As such, the effect of zero mode is negligible in a long chain.

Specifically, we consider a two-band model introduced in Ref.\cite{Song2019real}, with the Bloch Hamiltonian:
\bea
H(k)&=&d_x(k)\sigma_x+d_y(k)\sigma_y+i\kappa;\nn\\
d_x(k)&=&t_1+(t_2+t_3)\cos k+i\frac{\gamma}{2}\sin k\nn\\
d_y(k)&=&(t_1-t_3)\sin k+i\frac{\gamma}{2}\cos k. \label{twobandH}
\eea
The real-space OBC Hamiltonian is
\bea
H=\left(
\begin{array}{ccccc}
h_0 &  h_1 & 0 & 0 & \cdots\\
h_{-1} & h_0 & h_1 & 0 & \cdots\\
0 & h_{-1} & h_0 & h_1 & \cdots\\
0 & 0 & h_{-1} & h_0 & \cdots\\
\cdots & \cdots & \cdots & \cdots & \cdots
\end{array}
\right),
\eea
in which
\bea
h_0&=&t_1\sigma_x+i\kappa,\nn\\
h_{1}&=&\frac{t_2+t_3+\gamma/2}{2}\sigma_x+i\frac{\gamma/2-t_2+t_3}{2}\sigma_y,\nn\\
h_{-1}&=&\frac{t_2+t_3-\gamma/2}{2}\sigma_x+i\frac{\gamma/2+t_2-t_3}{2}\sigma_y.
\eea
We consider an OBC chain with $L$ unit cells, with each unit cell containing two sites $2j-1,2j$ ($j=1,\cdots,L$). The odd and even sites correspond to $\sigma_z=1$ and $-1$, respectively. To see the effects of zero modes, we calculate $G_{ij}$ for $i,j=1,2,2L-1,2L$ (Fig. \ref{zeromode}). A peak is found at $\omega=0$ for $G_{1,2L}$ and $G_{2L,1}$. This peak is not seen in $G_{2,2L-1}$ and $G_{2L-1,2}$, which is an evidence that it stems from the topological zero modes. In fact, each topological zero mode has a chirality, meaning that it is an eigenstate of $\sigma_z$. Specifically, the zero modes in our model, as eigenstates of $\sigma_z$, have vanishing weight at sites $j=2$ and $j=2L-1$. As such, $G_{2,2L-1}$ and $G_{2L-1,2}$ are insensitive to the zero modes, which explains the absence of zero-mode peak. A more significant feature is the length dependence of zero-mode peak. Comparing Fig. \ref{zeromode} (a) and (b), or (c) and (d), we see that the zero-mode peak diminishes as the chain length grows. This is intuitive because the effects of topological zero modes are expected to be significant only near the edges.

To compare the numerical results and our theory, we plot $\alpha_\rw$, $\alpha_\lw$ in Fig. \ref{twoband}, which are in excellent agreement with our formulas. Note that $h(\beta)$ is a matrix and $\beta_{j=1,2,3,4}$ are the roots of $\det[\omega-h(\beta)]=0$.

\section{Periodic-boundary condition}

In the main article, we focused on the OBC cases. This is not an arbitrary choice. In fact, it has been pointed out that when the OBC system has directional amplification, the corresponding PBC system tends to be dynamically unstable\cite{McDonald2018phase,Wanjura2019}. Moreover, it can be proved that even if dynamical stability is assumed, the PBC Green's function cannot have exponential growth with increasing spatial distance, and therefore they do not support directional amplification. Since GBZ reduces to the standard BZ in the PBC case, one can prove the absence of exponential growth based on BZ, following Ref. \cite{Wanjura2019}.

To prove the absence of exponential growth of $G_{ij}$ with $i-j$, let us consider a PBC chain of length $L$ with site coordinate $i=1,2,\cdots,L$. The Green's function reads
\bea
G^{\text{PBC}}_{ij}(\omega)&=&\frac{1}{L}\sum_k\frac{e^{ik(i-j)}}{\omega-h(k)},
\eea where $k=2\pi p/L$ with $p=0,1,2,\cdots,L-1$, as required by the PBC. Defining $\beta=\exp(ik)$, we can rewrite it as \bea
G^{\text{PBC}}_{ij}(\omega)&=&\frac{1}{L}\sum_\beta\frac{\beta^{i-j}}{\omega-h(\beta)}, \label{sum}
\eea where $\beta=\exp(2\pi i p/L)$, with $p=0,1,2,\cdots,L-1$. Now we can expand $\beta^{i-j}/(\omega-h(\beta))$ as a Laurent series at the BZ $|\beta|=1$:
\bea
\frac{\beta^{i-j}}{\omega-h(\beta)}&=&\sum_{n=-\infty}^{\infty} \beta^n \int_{|\tilde \beta|=1}\frac{d\tilde \beta}{2\pi i\tilde\beta}\frac{\tilde \beta^{i-j-n}}{\omega-h(\tilde \beta)}. \eea
We are then able to recast Eq. (\ref{sum}) into
\bea
G^{\text{PBC}}_{ij}(\omega) &=& \frac{1}{L}\sum_\beta \sum_{n=-\infty}^{\infty} \beta^n \int_{|\tilde \beta|=1}\frac{d\tilde \beta}{2\pi i\tilde\beta}\frac{\tilde \beta^{i-j-n}}{\omega-h(\tilde \beta)}. \quad \eea
Using the identity $ \frac{1}{L}\sum_\beta \beta^n  \equiv \frac{1}{L} \sum_{p=0}^{L-1} \exp(i\frac{2\pi p n}{L}) =\sum_m \delta_{n,mL}$, with $m$ integer-valued, we have
\bea
G_{ij}^{\text{PBC}}(\omega)=\sum_{m=-\infty}^{\infty}\int_{\text{BZ}}\frac{d\beta}{2\pi i \beta}\frac{\beta^{i-j-mL}}{\omega-h(\beta)}, \label{PBCformula}
\eea in which we have made the notational change of integration variable $\tilde{\beta}$ to $\beta$. The summation over $m$ guarantees that $G_{ij}=G_{i,j+L}$, as required by PBC, is satisfied. This expression is quite intuitive. Each $m$ value represents the times that a path winds around the chain, the sign of $m$ standing for the clockwise/anticlockwise sense. Intuitively, the contributions from larger $|m|$ is less significant, as the rapid phase oscillation of $\beta^{-mL}$ suppress the net contribution.

Taking advantage of the translational symmetry, we can fix $j=1$ and take $1\leq i\leq L$, then Eq. (\ref{PBCformula}) becomes
\bea
G_{i1}^{\text{PBC}}(\omega)=\sum_{m=-\infty}^{\infty}\int_{\text{BZ}}\frac{d\beta}{2\pi i \beta}\frac{\beta^{i-1-mL}}{\omega-h(\beta)}. \label{i1}
\eea
We order the roots of $\omega-h(\beta)=0$ as $|\beta_1|\dots\le|\beta_a|\le1\le|\beta_{a+1}|\dots\le|\beta_{2M}|$. It follows from the residue theorem that, for $m\leq 0$,
\bea
\int_{\text{BZ}}\frac{d\beta}{2\pi i \beta}\frac{\beta^{i-1-mL}}{\omega-h(\beta)}&=&c_1\beta_1^{i-1-mL}+\dots+c_a\beta_a^{i-1-mL}\nn\\
&\approx&c_a\beta_a^{i-1-mL},
\eea
and, for $m>0$,
\bea
\int_{\text{BZ}}\frac{d\beta}{2\pi i \beta}\frac{\beta^{i-1-mL}}{\omega-h(\beta)} &=& c_{a+1}\beta_{a+1}^{i-1-mL}+\dots+c_{2M}\beta_{2M}^{i-1-mL}\nn\\
&\approx&  c_{a+1}\beta_{a+1}^{i-1-mL},
\eea
where $c_1,\dots,c_{2M}$ are some numerical coefficients of order unity, and the ``$\approx$'' picks up the dominant term in the large $L$ limit.  Therefore, the $m$-summation in Eq. (\ref{i1}) reads
\bea
G_{i1}^{\text{PBC}}(\omega)\approx\sum_{m=-\infty}^{0}c_a\beta_a^{i-1-mL}+\sum_{m=1}^{\infty}c_{a+1}\beta_{a+1}^{i-1-mL}. \quad
\eea
For large $L$, the $m=0$ term dominates over all $m<0$ terms, while the $m=1$ term dominates all $m>1$ terms, and therefore we only retain these two terms:
\bea
G_{i1}^{\text{PBC}}(\omega)\approx  c_a\beta_a^{i-1}+c_{a+1}\beta_{a+1}^{i-1-L}. \label{twoterms}
\eea
Observing that this formula is obtained under the condition $1\leq i\leq L$, and that $|\beta_a|<1<|\beta_{a+1}|$, $i-1-L<0$, we see that $|\beta_a^{i-1}|<1$ and $|\beta_{a+1}^{i-1-L}|<1$.   Therefore, the PBC Green's function cannot have the exponential growth behavior found in the OBC chains.

\begin{figure}
\includegraphics[width=6cm, height=5.5cm]{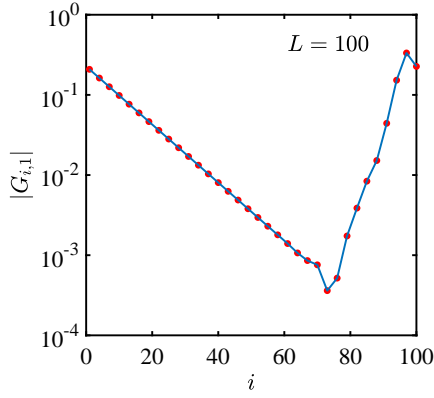}
\caption{ $|G_{i1}|$ of a PBC chain.  The dots are obtained from numerical real-space calculation. The solid curve represents the theoretical result from the BZ integral. The model is shown in the Fig. 1 of the main article, with $t_1=t_2=1$, $\gamma=4/3$, $\kappa=-0.8$, and $\omega=-1.7$. }
\label{Gi1}
\end{figure}

As an illustration, we show $|G_{i1}|$ for the model from the main article [Fig. \ref{Gi1}]. We find that $|G_{i1}|<1$ for all $i$, indicating the absence of exponential growth. This is consistent with the theory. Moreover, we find that the slope of the $|G_{i1}|$ curve changes at some point on the chain. This behavior can also be seen from the theory. In fact, when the first or second term in Eq. (\ref{twoterms}) dominates, we have
\bea G_{i1}^{\text{PBC}}(\omega)\approx  c_a\beta_a^{i-1}, \eea
or
\bea
G_{i1}^{\text{PBC}}(\omega)\approx   c_{a+1}\beta_{a+1}^{i-1-L},
\eea
respectively, and the transition point between these two regimes is given by their balance
\bea
|\beta_a^{i_0-1}|=|\beta_{a+1}^{i_0-1-L}|,
\eea
which leads to $i_0=\frac{\log|\beta_{a+1}|}{\log|\beta_{a+1}/\beta_a|}L+1$. For the parameter values in Fig. \ref{Gi1}, we have $i_0\approx 77$, which is in accordance with the numerical value.

\section{\label{sec:level3}Derivation of the effective non-Hermitian Hamiltonian from the quantum master equation}

By the definition of $\psi_i$, we have
\bea
\dot \psi_i = \frac{d}{dt} \text{Tr}[a_i\rho(t)] =\text{Tr}[a_i\dot\rho(t)].
\eea
Inserting the master equation
\bea
\dot\rho(t)=-i[H_0,\rho]+\sum_\mu\left(L_\mu\rho L_\mu^\dagger-\frac{1}{2}\{L_\mu^\dagger L_\mu,\rho\}\right),
\eea
we obtain
\bea
\dot{\psi}_i &=& -i\text{Tr} ( [a_i,H_0]\rho(t) ) \nn\\  &&+\frac{1}{2}\sum_\mu   \text{Tr}\left([L_\mu^\dagger,a_i]L_\mu\rho(t)
 + L_\mu^\dagger[a_i,L_\mu]\rho(t) \right). \label{three}
\eea
Now we take the Hamiltonian
\bea
H_0&=&\sum_{i,j}(h_0)_{ij}a_i^\dagger a_j + \sum_i(\epsilon_i a_i^\dag + \epsilon^*_i a_i), \eea
and the dissipators \bea \{L_\mu\} = \{L_i^l,\, L_i^g\}, \eea with
\bea
L_i^l&=&\sum_jD_{ij}^la_j; \quad L_i^g=\sum_jD_{ij}^ga_j^\dagger.
\eea
The specific system shown in Fig. 1(d) of the main article is a special case of this general model.
The first term in Eq. (\ref{three}) is
\be
-i\text{Tr}\left([a_i,H_0]\rho(t)\right)=-i\sum_j (h_0)_{ij}\psi_j -i\epsilon_i(t).
\ee
For the second term in Eq. (\ref{three}), only the loss dissipators $L_i^l$ contribute:
\bea
\sum_m[L_m^{l\dagger},a_i]L_m^l &=&\sum_{m,n}D_{mn}^{l*}[a_n^\dagger,a_i]L_m^l\nn\\
&=&-\sum_mD_{mi}^{l*}L_m^l\nn\\
&=&-\sum_{m,j}D_{mi}^{l*}D_{mj}^la_j,
\eea
while for the third term, only the gain dissipators $L_i^g$ contribute:
\bea
\sum_mL_m^{g\dagger}[a_i,L_m^g]&=&\sum_{m,n}L_m^{g\dagger}[a_i,D_{mn}^ga_n^\dagger]\nn\\
&=&\sum_m D_{mi}^g L_m^{g\dagger}\nn\\
&=&\sum_{m,j}D_{mi}^g D_{mj}^{g*} a_j.
\eea
Thus, the second and third terms of Eq. (\ref{three}) are simplified to
\bea \frac{1}{2}\left( (D^{g\dag}D^g)^T - D^{l\dag}D^l \right)_{ij}\psi_j. \eea
Summing up these terms, Eq. (\ref{three}) becomes \bea \dot{\psi}_i=-i\sum_j H_{ij}\psi_j -i\epsilon_i(t), \eea with the effective non-Hermitian Hamiltonian
\bea H= h_0 + \frac{i}{2}\left( (D^{g\dag}D^g)^T - D^{l\dag}D^l \right). \label{Hgeneral} \eea
For our specific model, the nonzero parameters are $(h_0)_{i,i+1}=(h_0)_{i+1,i}=t_1$, $(h_0)_{i,i+2}=(h_0)_{i+2,i}=t_2$, $(h_0)_{ii}=\omega_0$, $D^l_{i,i}=\sqrt{\gamma}$, $D^l_{i,i+2}=-i\sqrt{\gamma}$ and $D^g_{i,i}=\sqrt{\gamma'}$. It follows from Eq. (\ref{Hgeneral}) that the effective non-Hermitian Hamiltonian $H$ is given by Fig. 1(a) of the main article. We note that under the OBC, while diagonal elements in the bulk are all $\omega_0+i\kappa=\omega_0+i(\gamma'/2-\gamma)$, the four edge-site diagonal elements $H_{11},H_{22},H_{L-1,L-1},H_{LL}=\omega_0+i(\gamma'-\gamma)/2$.  For simplicity, we let these four elements be $\omega_0+i\kappa$, which does not cause any appreciable modification of our main results (While this is intuitively apparent, we have also numerically confirmed it). The $H$ operator reads
\bea
H=\left(
\begin{array}{ccccc}
\omega_0+i\kappa &  t_1 & t_2-\frac{\gamma}{2} & 0 & \cdots\\
t_1 & \omega_0+i\kappa  & t_1 & t_2-\frac{\gamma}{2} & \cdots\\
t_2+\frac{\gamma}{2} & t_1 & \omega_0+i\kappa & t_1 & \cdots\\
0 & t_2+\frac{\gamma}{2} & t_1 & \omega_0+i\kappa  & \cdots\\
\cdots & \cdots & \cdots & \cdots & \cdots
\end{array}
\right). \eea
To simplify the expressions, we measure the frequency with respect to $\omega_0$, i.e., shift the frequency (energy) $\omega\rw\omega-\omega_0$:
\bea
H=\left(
\begin{array}{ccccc}
 i\kappa &  t_1 & t_2-\frac{\gamma}{2} & 0 & \cdots\\
t_1 & i\kappa  & t_1 & t_2-\frac{\gamma}{2} & \cdots\\
t_2+\frac{\gamma}{2} & t_1 & i\kappa & t_1 & \cdots\\
0 & t_2+\frac{\gamma}{2} & t_1 &  i\kappa  & \cdots\\
\cdots & \cdots & \cdots & \cdots & \cdots
\end{array}
\right). \eea
We could also include more dissipators such as
$\sqrt{\gamma_1}(a_j-ia_{j+2}),  \sqrt{\gamma_2}(a_j+ia_{j+2}),
\sqrt{\gamma_3}(a_j^\dagger+ia_{j+2}^\dagger), \sqrt{\gamma_4}(a_j^\dagger -ia_{j+2}^\dagger),
\sqrt{\gamma'_1}a_j^\dagger, \sqrt{\gamma'_2}a_j$, and it is straightforward to obtain a similar $H$, except that the diagonal elements become $H_{ii}=(\gamma'_1-\gamma'_2)/2-\gamma_1-\gamma_2+\gamma_3+\gamma_4$, and $H_{i,i\pm 2}$ become $t_2\pm(\gamma_2+\gamma_3-\gamma_1-\gamma_4)/2$. Our Fig. 1 in the main article corresponds to the special case $\gamma_1=\gamma, \gamma'_1=\gamma'$ with other $\gamma$'s vanishing.

\end{document}